\documentclass[a4paper,onecolumn,11pt,accepted=2026-01-26]{quantumarticle}
\pdfoutput=1
\usepackage[utf8]{inputenc}
\usepackage[english]{babel}
\usepackage[T1]{fontenc}
\usepackage{adjustbox}
\usepackage{algorithm}
\usepackage[noend]{algpseudocode}
\usepackage{amsmath}
\usepackage{amssymb}
\usepackage{amsthm}
\usepackage{booktabs}
\usepackage{enumitem}
\usepackage{float}
\usepackage[citecolor=blue]{hyperref}
\usepackage{mathrsfs}
\usepackage{mathtools}
\usepackage{multirow}
\usepackage{physics}
\usepackage[justification=centering]{subcaption}

\usepackage{tikz}
\usetikzlibrary{quantikz2}
\usepackage{lipsum}
\usepackage{pgfplots}

\newtheorem{theorem}{Theorem}

\newtheorem{proposition}{Proposition}
\newtheorem{corollary}{Corollary}

\theoremstyle{definition}
\newtheorem{definition}{Definition}
\newtheorem{example}{Example}
\theoremstyle{remark}
\newtheorem{remark}{Remark}[section]

\AtBeginEnvironment{example}{%
  \pushQED{\qed}%
}
\AtEndEnvironment{example}{\popQED\endexample}

\newcommand{\twodots}{\mathinner {\ldotp \ldotp}}
\newcommand{\intset}[2]{\ensuremath{\left[#1\twodots{} #2\right]}}

\newcommand{\pres}[2]{{\langle\!\hskip1pt\!\langle #1 | #2 \rangle\!\hskip1pt\!\rangle}}

\DeclareMathOperator{\vecop}{\mathrm{vec}_r}		

\DeclareRobustCommand
  \compactcdots{\mathinner{\cdotp\mkern-2mu\cdotp\mkern-2mu\cdotp}}
  
\newcommand{\Algo}{\textsc{QuPer}}
\newcommand{\QAOA}{\textsc{Qaoa}}
\newcommand{\ADAM}{\textsc{Adam}}

\begin{document}

\title{Variational quantum algorithms for permutation-based combinatorial problems:

Optimal ansatz generation with applications to quadratic assignment problems and beyond}

\author{Dylan Laplace Mermoud}
\affiliation{UMA, ENSTA, Institut Polytechnique de Paris, 91120 Palaiseau, France}
\affiliation{CEDRIC, Conservatoire National des Arts et M{\'e}tiers, 75003 Paris, France}
\orcid{0000-0002-1508-9656}
\email{dylan.laplace@ensta.fr}
\thanks{This research benefited from the support of the FMJH Program PGMO, project number 2023-0009, and from the ANR project HQI-ANR-22-PNCQ-0002.}
\author{Andrea Simonetto}
\affiliation{UMA, ENSTA, Institut Polytechnique de Paris, 91120 Palaiseau, France}
\affiliation{CEDRIC, Conservatoire National des Arts et M{\'e}tiers, 75003 Paris, France}
\orcid{0000-0003-2923-3361}
\author{Sourour Elloumi}
\affiliation{UMA, ENSTA, Institut Polytechnique de Paris, 91120 Palaiseau, France}
\affiliation{CEDRIC, Conservatoire National des Arts et M{\'e}tiers, 75003 Paris, France}
\affiliation{ensIIE, École Nationale Supérieure d'Informatique pour l'Industrie et l'Entreprise, 91025, Évry, France}
\orcid{0000-0001-6289-7958}
\maketitle


\begin{abstract}
\noindent  {\bf Abstract.} We present a quantum variational algorithm based on a novel circuit that generates all permutations that can be spanned by one- and two-qubits permutation gates. The construction of the circuits follows from group-theoretical results, most importantly the Bruhat decomposition of the group generated by the \(\mathtt{cx}\) gates. These circuits require a number of qubits that scale logarithmically with the permutation dimension, and are therefore employable in near-term applications. We further augment the circuits with ancilla qubits to enlarge their span, and with these we build ansatze to tackle permutation-based optimization problems such as quadratic assignment problems, and graph isomorphisms. The resulting quantum algorithm, \Algo{}, is competitive with respect to classical heuristics and we could simulate its behavior up to a problem with $256$ variables, requiring $20$ qubits. 
\end{abstract}

\vfill

\tableofcontents

\vfill

\newpage

\section{Introduction}

Combinatorial optimization has been advocated as one of the key benchmarks for quantum computing: its classically hard nature combined with its wide societal importance makes it an attractive area of research for quantum algorithms. One of such algorithms is the celebrated quantum approximate optimization algorithm, or  \QAOA{}, specifically designed for noisy intermediate-scale quantum computers (or NISQ)~\cite{Farhi2014,Zhou2020}, which now exists in many variants, see~\cite{Abbas2024} for an account. Other algorithms have been also proposed in other noise regimes, but they will not be our main focus in this paper. 

Even though  \QAOA{} has been widely studied, its performance on hard combinatorial problems is still under scrutiny. Theoretically, we know that  \QAOA{} is not classically simulable, and it may have some, albeit limited, advantage with respect to some classical algorithm~\cite{Boulebnane2024,Shaydulin2024,Aaronson2024}; however, more broadly,  \QAOA{} is not readily expected to solve relevant combinatorial problems at scale. There are at least two main issues that would challenge  \QAOA{} to achieve that. 

First, despite their hardness, there exist methods performing well at solving combinatorial problems classically, or at least at finding approximate solutions. As Berstimas and Dunn describe in a recent paper, \emph{``in the last 25 years, algorithmic advances in integer optimization coupled with hardware improvements have resulted in an astonishing 800 billion factor speedup in mixed-integer optimization''} (a class of hard combinatorial problems)~\cite{Bertsimas2017}. This translates, for example, into solving some sparse instances of quadratic unconstrained binary optimization problems (QUBOs) exactly up to tens of thousands of variables~\cite{Rehfeldt2023}.  \QAOA{} would require at least tens of thousands of logical qubits to achieve a similar feat.   

Second, in general, \QAOA{} cannot handle constraints in an hard form. We are forced to soft-constrain the problem by lifting the constrains into the cost, thereby creating numerical issues and approximations. Constraints are at the heart of optimization: we model and then simplify optimization problems by adding constraints of all sorts.
 
In order to overcome the first issue, mainly due to the basis encoding of the search space in  \QAOA{}, we have seen some interest in amplitude encodings~\cite{Weigold2021,GonzalezConde2024}. Amplitude encodings require a number of qubits that scales logarithmically with the number of the variables of the optimization problem, thereby being more resource-friendly. Amplitude encodings present other issues, however. The fact that they encode directly vector components and matrices yield typically non-diagonal observables. Furthermore, in most cases, they are trivially classically simulable. Take for example the max-cut formulation in~\cite{Rancic2023}. If $n$ indicates the number of decision variables of a max-cut formulation, then the approach requires only $O(\log n)$ qubit to run. However, the computation of the observable, i.e., the cost, requires $O(n^2)$ calls to the quantum circuit and a total number of $O(n^3)$ floating operations (cf.~Appendix~\ref{ap.1}). To run the same heuristic classically, without quantum computations, we would only require $O(n^2)$ floating point operations to evaluate the cost.  

In contrast, in~\cite{Mariella2023} the authors propose an amplitude encoding of the subgraph isomorphism problem. This is a genuine quantum algorithm, requiring $O(\log(n))$ qubits to represent a problem in $n$ variables, and with a circuit delivering directly the cost function (the observable is the projector on the first basis state), so a simple average of a number of samples is required classically to retrieve the expected value of the quantum cost. In contrast, a solely classical approach would require $O(n^2)$ floating point to compute the cost.  So dealing with amplitude encodings requires great care.

In order to tackle the second issue (the constraints), different variations of \QAOA{} have been proposed, in which we encode in one way or another some constraints into the variational circuit to optimize over. A popular approach is to encode the constraints into the mixer Hamiltonian~\cite{Wang2020, Fuchs2022}, which works well for some constraints but might be hard in general~\cite{leipold2024imposing}. Another approach is to map the feasible space and define a continuous quantum walk that transverses this space~\cite{Marsh2020}. Other approaches are presented in~\cite{Gambella2020,DriebSchoen2023,Nakada2025}. All these approaches are constraint-dependent and either hard to generalize, to implement, or their accuracy is limited. For example, mapping the whole feasible space classically may even be as hard as solving the original combinatorial problem. 

\smallskip

With this in mind, in this paper, we depart from standard \QAOA{} with the aim to develop variational quantum algorithms that can tackle one typical combinatorial constraint at scale. We focus here on permutation-based binary optimization problems~\cite{Ceberio2012}: problems which are defined over permutations, which are widespread in operations research. For example, the quadratic assignment problem, the graph isomorphism problem, and the travel salesperson problem are permutation-based optimization problems. We argue that these problems provide a very useful baseline to construct more complicated problems. 

Besides their usefulness, we focus here on permutations, since they have quite a natural quantum structure. Permutations are in fact unitary matrices and they can be encoded in quantum gates natively. Furthermore, we will see that permutations have a number of useful mathematical properties that makes them the ideal candidate for quantum circuits. 

To try to overcome some of the drawbacks of \QAOA{}, instead of interpreting the cost as observable or encoding it in a quantum circuit, we encode directly a permutation ansatz. The circuit, when measured, directly delivers a doubly-stochastic matrix, i.e., a positive superposition of permutation matrices. This has several advantages: first, the cost can be computed classically and can be anything. This renders the approach extremely general, which allows us to test it on various problems. Second, we can use amplitude encodings to reduce the number of used qubits without having the problem of a non-diagonal observable. This win-win setting is also made non classically simulable by adding additional ancilla qubits, following the very clever circuit design of Mariella and coauthors~\cite{Mariella2024}. 

More specifically, in this paper, we offer the following contributions.
\begin{enumerate}[topsep=3pt, itemsep=-1pt]
\item We study variational quantum circuits that can span a given permutation set. In particular, we construct a quantum circuit made of rotations and parametrized \texttt{cx} gates that can deliver \emph{all} the permutations available by using these local gates alone. For a number of qubits $q = \log_2 n$ if $n$ is the permutation size, the circuit has $O(q^2)$ classical parameters, $O(q)$ depth, and spans $2^{O(q^2)}$ permutations.  

\item We propose a quantum circuit that can bridge the gap between the maximum span and the total number of permutations, i.e., $2^{O(q^2)}$ vs. $(2^q)!$, by using ancilla qubits. We study how the ancilla qubits can deliver doubly-stochastic matrices, their maximum span, and how to construct permutations with them.  

\item We test the circuit on several optimization problems of wide-reaching interest proposing \Algo{}, a variational quantum algorithm for permutation-based optimization problems. In particular, we tackle the quadratic assignment problem and the graph isomorphism. We show that we are competitive with classical heuristic on small and sometimes bigger instances with $n=256$ variables, and we perform better than other variational quantum algorithms, when they exist. 
 
\end{enumerate}

\subsection{Closely related work}

Classically, permutation-based problems have received a lot of attention. We will report in the sections related to each problem that we simulate an account of the literature for that specific problem. However, we can already refer to~\cite{Fogel2015} as a typical convex approach for the this general class of optimization problems. This approach would convexify the constraints, by letting the permutation being a doubly-stochastic matrix, solve the problem exactly, and then project the solution onto the permutation set. This particular approach can offer good approximate solutions when the cost is convex in the permutation variable. Our quantum approach is somewhat similar, since it generates doubly-stochastic matrices, but it is a non-convex heuristic, instead of being convex, and it can be applied to a large variety of problems. Of~\cite{Fogel2015}, we keep the projection-onto-permutations strategy~\cite{Barvinok2006}, which is standard in classical optimization and which we will describe in details.

The permutation circuit construction that we propose is largely inspired from the circuit of~\cite{Mariella2024}. There, Mariella and coauthors propose a non-classically simulable circuit to generate doubly-stochastic matrices that they then use to solve an optimal transport problem. In this paper, we study with great care which unitary to use within the circuit to span as much as possible the permutation set, and we prove additional results on the span of the obtained double-stochastic matrices. We see that this is not a trivial question and different choices of the unitary lead to very different results. 

In order for us to study the span of certain unitaries that we construct based on rotation and parametrized \texttt{cx} gates, we make heavy use of group theory concepts, such as semi-direct products and the Bruhat decomposition. Here, we build on the work of~\cite{bataille2022quantum}, who studied quantum circuits of \texttt{cx} gates and their span. But then we largely generalize his work to rotation gates, parametrized \texttt{cx}s, and their combinations. We will provide specific references as we go along.    
As a side note, the Bruhat decomposition has been also used in the context of stabilizer circuits, where the decomposition of the symplectic groups allows for designed circuits of reduced depth~\cite{maslov2018shorter}.

\section{General problem formulation and main results}\label{sec:pf}

We are interested in solving the problem
\begin{equation}\label{eq:perm:prob}
\min_{P \in \Pi_n} \, f(P),
\end{equation}
where $\Pi_n$ is the set of permutation matrices of dimension $n\times n$, and $f:\Pi_n \to \mathbb{R}$ is a cost function to minimize. We let $n = 2^q$ for any integer $q$, without loss of generality. Problem~\eqref{eq:perm:prob} is binary, combinatorial, and in general \textsf{NP}-Hard. When $f$ is linear in $P$, then the problem reduces to the linear assignment problem, a convex optimization problem that is easily solvable at scale, e.g., with the Auction algorithm~\cite{Bertsekas1992}. In most of the other cases, Problem~\eqref{eq:perm:prob} is very hard to solve classically even for $n$ in the order of a few hundreds. In fact, the feasible space for~\eqref{eq:perm:prob}, that is the number of permutations of a given order $n$, is $n!$, which grows faster than an exponential.   

\subsection{Spanning permutations via a parametric circuit}

To tackle Problem~\eqref{eq:perm:prob}, we turn to variational quantum algorithms. In particular, we will devise a parametric quantum circuit $\mathscr{P}(\theta)$ that will be able to represent a large subset of the whole permutation set. When measuring the circuit, we will obtain the coefficients of a doubly-stochastic matrix $\hat{P}_\theta$, that can then be used classically to optimize a certain cost $f(\hat{P}_{\theta})$. On the classical side, we will be using a classical optimizer (hereunder \ADAM{}~\cite{Kingma2014}) to optimize for $\theta$, while asking to the quantum circuit the value of $\hat{P}_{\theta}$ for each given $\theta$. During the run, we have the possibility to classically project the obtained doubly-stochastic matrix $\hat{P}_\theta$ onto the permutation set, thereby obtaining an approximate solution $\tilde{P}$. 

The advantage of the approach proposed in this paper lies in the representation of the quantum circuit $\mathscr{P}(\theta)$, which allows us to span a larger set of the permutation set with a limited number of continuous parameters $\theta$. We will see that this circuit is, in general, non classically simulable, thereby granting our variational quantum algorithm the label of a genuine quantum approach. 

\begin{figure}[H]
\centering
\includegraphics[width=0.8\textwidth]{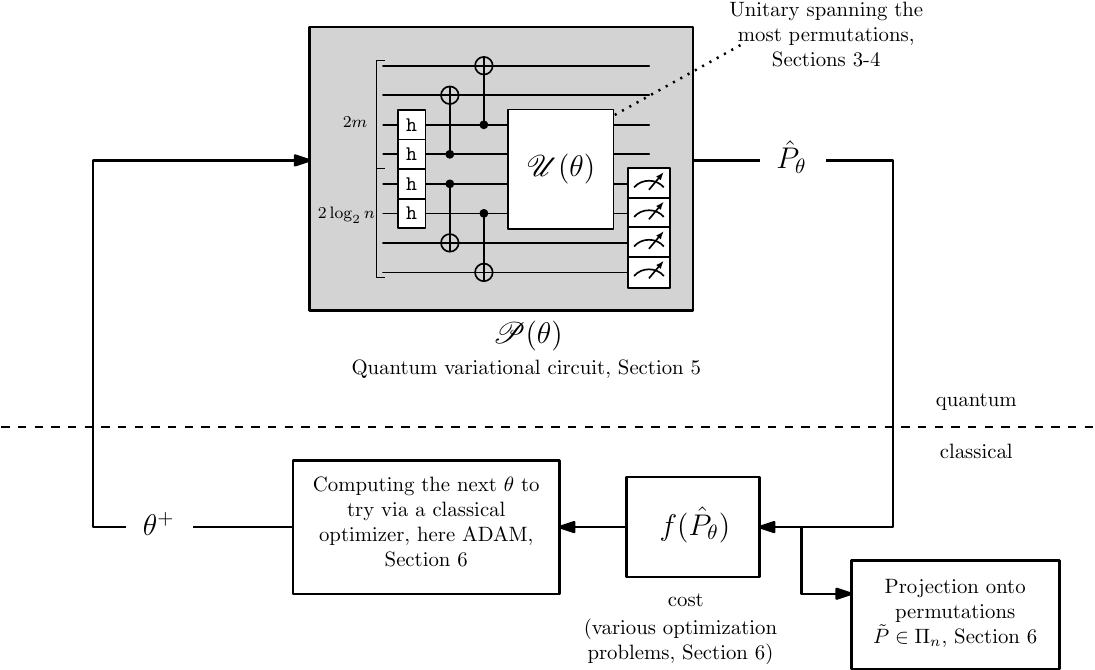}
\caption{Pictorial description of the proposed variational approach, labeled \Algo{}, with the reference to all the sections where each element is discussed. }
\label{fig.approach}
\end{figure}

We report in Figure~\ref{fig.approach}, the aforementioned approach. We remark that $\mathscr{P}(\theta)$ is designed as such that permutations are encoded in amplitudes, so that the number of required qubits will scale only logarithmically with the permutation set $n$. $\mathscr{P}(\theta)$ will be also augmented by an arbitrary even number $2m$ of ancilla qubits to guarantee both a larger span and a non-trivially classically simulable circuit. 

The first important result that we will discuss in great detail pertains how much of the permutation set we can span with only rotation and parametrized \texttt{cx} gates and how to construct such a circuit. These gates are the basic gates that we will use in our construction, since readily available in most quantum hardware. The result is given informally below. 

\begin{theorem}[Informal]\label{th.1.informal} Consider a $q$-qubit circuit whose unitary is constructed to represent a $n$-dimensional permutation. 
\begin{enumerate}[topsep=3pt, itemsep=-1pt, label=(\roman*)]
\item The permutations spanned only by using rotation and parametrized \(\mathtt{cx}\) gates can be computed by considering the composition of two circuits: one containing rotations only and one containing parametrized \(\mathtt{cx}\) gates only. 
\item Any circuit containing parametrized \(\mathtt{cx}\) gates only can be decomposed into three circuits, by the Bruhat decomposition. 
\item The circuit spanning all the permutations available by considering only rotations and parametrized \(\mathtt{cx}\) has $O(q^2)$ parameters, $O(q)$ depth, and spans $2^{O(q^2)}$ permutations. 
\end{enumerate}
\end{theorem}

Theorem~\ref{th.1.informal} will be made formal in Sections~\ref{sec:prelims} and \ref{sec:circuits}, and specifically by Propositions~\ref{prop: presentation-lx} and \ref{prop: semi-direct-product} as well as Theorems~\ref{th: decomposition},~\ref{th: transvection-generation}, and~\ref{th:number-of-permutations}. Theorem~\ref{th.1.informal} informs us that to obtain the maximum span that local gates have to offer, we only need a number of classical parameters that scales quadratically in the number of qubits and a linear depth. We will construct such circuit, see for instance Figure~\ref{fig:ansatze}. 

Naturally, $2^{O(q^2)}$ is lower than the total number of permutations $(2^q)!=n!$. The second important result is how to bridge this gap by using $m$ ancilla qubits. We will make use of the very clever circuit design of Mariella and coauthors~\cite{Mariella2024}, to generate doubly-stochastic matrices. We will present such circuit in Figure~\ref{fig:mariella} and we will study in details the choice of ansatze to use. In addition, we will be able to show that with our ansatze for permutations, we can span at most $2^{O((q+m)^2)}$ permutations, and therefore we will need at least $m \gtrsim \sqrt{n \log n}$ ancilla to span the whole permutation set of $n!$. We will make this formal in Proposition~\ref{prop.decomps}.

\subsection{Solving optimization problems}

Given the circuit and the ansatze that we have designed, we will then set forth to solve permutation-based optimization problems, such as the quadratic assignment problem and the graph isomorphism. We will propose a novel variational algorithm, labelled \Algo{}, which we will show competitive with respect to classical approaches, at least in small instances but not only, and with which we will be able to simulate problems up to $n=256$ variables in $20$ qubits. 

\subsection{Organization of the paper}

The remainder of the paper is organized as follows. In Sections~\ref{sec:prelims} and \ref{sec:circuits}, we discuss how to construct quantum circuits representing permutations: these are the most mathematically heavy sections. We then propose the variational circuit $\mathscr{P}(\theta)$ in Section~\ref{sec:p-circuit} and analyze its theoretical and numerical properties. In Section~\ref{sec:optimiz}, we present numerical results for two optimization problems of widespread interest: the quadratic assignment problem and the graph isomorphism. We then conclude.

\section{Spanning permutations}\label{sec:prelims}

\subsection{Preliminaries}

The first step of our variational quantum circuit construct is a mathematical one. Given simple local gates, hereunder \(\mathtt{x}\) and \(\mathtt{cx}\), how many distinct permutation matrices can we construct? We will see how this is important in our variational circuit in Section~\ref{sec:p-circuit}. 

As such, we start by defining a quantum circuit acting on $q$ qubits, and the  
the gates \(\mathtt{x}\) and \(\mathtt{cx}\), defined by 
\[
\mathtt{x}_k = \dyad{0}{1}_k + \dyad{1}{0}_k \qquad \text{and} \qquad \mathtt{cx}_{kl} = \dyad{0}{0}_k \otimes \mathtt{id}_l + \dyad{1}{1}_k \otimes \mathtt{x}_l. 
\]
The first gate \(\mathtt{x}_k\) acts on the single qubit $k$, while the second gate \(\mathtt{cx}_{kl}\) acts on two qubits, $k$ and $l$, respectively. Hence, \(\mathtt{x}\) exchanges the two amplitudes of a one-qubit state, and \(\mathtt{cx}\) exchanges the amplitudes of the state of the \(l\)th qubit if the state of the \(k\)th qubit is \(\ket{1}\). These two gates therefore act as permutation matrices on the space \(\left( \mathbb{C}^2\right)^{\otimes q}\) of states of \(q\)-qubit systems.

We need now to introduce some useful group theory concepts that will help us building and counting distinct permutations. The reader is also referred to~\cite{dummit2004abstract, johnson1997presentations} for good introductions on the subject. 

For any group, one can define the generators of the group, as a subset of elements of the group that can generate the whole group. More formally, a set of \emph{generators} of a given group \(G\) is a subset \(S \subseteq G\) such that any element of the group can be written as a word of generators, and their inverses. Together with a set of \emph{relations} \(R\) that the elements of the group, written as words of generators, satisfy, the set of generators characterizes the group \(G\). The pair $\pres{S}{R}$ is called a \emph{presentation}\footnote{We use here the notation $\pres{\cdot}{\cdot}$ instead of the more common $\langle\cdot|\cdot\rangle$ to avoid confusion with the inner product of two quantum states.} of the group \(G\). 

The group of all bijections of \(\{1, \ldots, n\}\), equipped with the law of composition, is called the \emph{symmetric group}, here written as \(S_n\). The elements of \(S_n\) are called \emph{permutations}. A well-known presentation for the symmetric group is the one given by \emph{adjacent transpositions}: i.e., any permutation can be generated by permuting (swapping) adjacent elements. Since we will make use of this, we describe it in details in the following example.

\begin{example}[Symmetric group \(S_n\)]\label{example.1}
A well-known presentation for the symmetric group is the one given by adjacent transpositions. Indeed, any permutation can be written as a word of adjacent transpositions. For instance, the permutation \((3 \, 2 \, 1)\)\footnote{We used here the \emph{one-line} notation for permutations. The permutation \((\sigma_1 \compactcdots \sigma_i \compactcdots \sigma_n)\) sends \(i\) to \(\sigma_i\).} which reverses the order of three elements, can be written as 
\[
(3 \, 2 \, 1) = (2 \, 1 \, 3) (1 \, 3 \, 2) (2 \, 1 \, 3). 
\]
Usually, the transposition \((\compactcdots \, k+1 \;  k \, \compactcdots)\) is denoted by \(\sigma_k\). The set of relations defining the symmetric group from these generators are, for all \(1 \leq k, j \leq n-1\) such that \(j \neq k \pm 1\), 
\[
\sigma_k^2 = \mathrm{id}, \qquad \sigma_k \sigma_j = \sigma_j \sigma_k, \qquad \text{and} \qquad \sigma_k \sigma_{k+1} \sigma_k = \sigma_{k+1} \sigma_k \sigma_{k+1}. 
\]
The last set of relations can be transformed into \((\sigma_k \sigma_{k+1})^3 = \mathrm{id}\) using \(\sigma_k^2 = \mathrm{id}\), where $\mathrm{id}$ represents the identity. 
\end{example}
 
The language of group theory using presentations can be directly transposed in the language of quantum circuits, at the very least for three reasons. First, a quantum circuit can be interpreted as a word of tensor products of basic gates. Secondly, the concept of generators is tightly connected to the one of basic gates: in order to write efficient circuits for the group under consideration, the generators must be written efficiently, as short words of basic gates. Finally, the relations give rewriting rules to design shorter circuits.

We consider now the group generated by \texttt{x} gates alone.

\subsection{The group generated by \texttt{x} gates}

Let \(X_q\) denote the group generated by \(\{ \mathtt{x}_k \mid 0 \leq k \leq q-1\}\). The group is abelian (i.e., any pair of elements commutes), and contains \(2^q\) elements. Because each generator only applies to one qubit and is an involution, any element of the group can be seen as a tuple of generators, exhibiting the direct sum structure of the group:
\[
X_q \simeq \bigoplus_{k = 0}^{q-1} \mathbb{Z}_2 =: \mathbb{Z}_2^q. 
\]
Here, \(\mathbb{Z}_2\) denotes the group defined on the set \(\{0, 1\}\) equipped with the bitwise `exclusive-or' additive law \(\oplus\) defined by 
\[
1 \oplus 1 = 0 \oplus 0 = 0, \qquad \text{and} \qquad 1 \oplus 0 = 0 \oplus 1 = 1.
\]
A presentation for \(X_q\) is therefore given by
\[
\pres{\mathtt{x}_0, \ldots, \mathtt{x}_{q-1} \, \,}{\, \mathtt{x}_k^2 = \mathtt{id} \text{ and } \left(\mathtt{x}_k \mathtt{x}_l\right)^2 = \mathtt{id}, \text{ for all } 0 \leq k, l \leq q-1}. 
\]
It is customary in group theory to write the relators as words of generators that correspond to the identity. However, this expression can be rewritten as follows. We can apply the exponentiation to get \(\mathtt{x}_k \mathtt{x}_l \mathtt{x}_k \mathtt{x}_l = \mathtt{id}\). By multiplying on both sides by \(\mathtt{x}_l \mathtt{x}_k\), and remembering that each \(\mathtt{x}\) is an involution, i.e., its own inverse, we obtain \(\mathtt{x}_k \mathtt{x}_l = \mathtt{x}_l \mathtt{x}_k\), so \(\mathtt{x}_k\) and \(\mathtt{x}_l\) commute.

\subsection{The group generated by \texttt{cx} gates}

Let \(CX_q\) denote the group generated by \(\{\mathtt{cx}_{kl} \mid k \neq l, 0 \leq k, l \leq q-1\}\). This group is no longer abelian, because we have
\[
\mathtt{cx}_{kl} \mathtt{cx}_{lm} \neq \mathtt{cx}_{lm} \mathtt{cx}_{kl}, \qquad \text{for } k, l, m \text{ all distinct}. 
\]
A presentation of \(CX_q\) was identified by Bataille~\cite{bataille2022quantum}. 

\begin{proposition}[\cite{bataille2022quantum}]
A presentation for the group \(CX_q\) is given by the set of generators \(\{\mathtt{cx}_{kl} \mid k \neq l, 0 \leq k, l \leq q-1\}\) and the relations
\begin{itemize}
\setlength\itemsep{0.5pt}
\item involution: \(\mathtt{cx}_{kl}^2 = \mathtt{id}\) where \(k \neq l\), 
\item commutation: \((\mathtt{cx}_{kl} \mathtt{cx}_{mp})^2 = \mathtt{id}\) where \(k \neq p\) and \(l \neq m\),
\item non-commutation: \((\mathtt{cx}_{kl} \mathtt{cx}_{lm})^2 = \mathtt{cx}_{km}\) where \(k, l, m\) are distinct. 
\end{itemize}
\end{proposition}

Let \(\mathrm{GL}_n(\mathbb{F}_2)\) be the group of invertible \((n \times n)\)-matrices over the field \(\mathbb{F}_2\). Steinberg~\cite{steinberg2016lectures} gave a presentation of it that perfectly coincides with the one of \(CX\) given by Bataille~\cite{bataille2022quantum}.

\begin{proposition}[\cite{steinberg2016lectures, bataille2022quantum}]\label{prop: isomorphism-linear}
The group \(CX_q\) is isomorphic to \(\mathrm{GL}_q(\mathbb{F}_2)\). 
\end{proposition}

This result will allow us to leverage the properties of \(\mathrm{GL}_q(\mathbb{F}_2)\), to implement the full \(CX_q\) in quantum circuits. In particular, Bruhat and Tits~\cite{bruhat1972groupes} demonstrated that any element of \(\mathrm{GL}_q(\mathbb{F}_2)\) can be written as the product of three elements, belonging to specific subgroups of \(\mathrm{GL}_q(\mathbb{F}_2)\). Let us see how. 

\medskip 

Let \(B\) denote the subgroup of \(\mathrm{GL}_q(\mathbb{F}_2)\) only composed of upper-triangular matrices, and let \(W\) denote the subgroup only composed of permutation matrices. 

\begin{theorem}[The Bruhat decomposition~\cite{bruhat1972groupes, bourbaki1968groupes}]\label{th: bruhat-decomposition}
\(\mathrm{GL}_q(\mathbb{F}_2)\) is isomorphic to \(B W \hspace{-1.5pt} B\). 
\end{theorem}

The Bruhat decomposition will be the cornerstone of our construction. It allows us to decompose the circuit synthesis problem into two, much simpler, circuit synthesis problems.

\subsection{The group generated both by \texttt{x} and \texttt{cx} gates}

In this subsection, we present \emph{new results} about the structure of the whole group generated by \(\mathtt{x}\) and \(\mathtt{cx}\) gates, and explain how it relates to the two groups described earlier. This group includes all the circuits on \(q\) qubits composed solely of \(\mathtt{x}\) and \(\mathtt{cx}\) gates, up to rewriting rules, i.e., relations, that we present in the following. 

Let \(LX_q\) denote the group generated by 
\[
\{ \mathtt{x}_k \mid 0 \leq k \leq q-1 \} \cup \{ \mathtt{cx}_{kl} \mid k \neq l, 0 \leq k, l \leq q-1 \},
\]
the set of all local permutation gates. Necessarily, it contains both \(X_q\) and \(CX_q\) as a subgroup, and therefore is subject to the same relations as they do. However, new elements generated both by \(\mathtt{x}\) and \(\mathtt{cx}\) are generated, which are subject to new relations. For instance, we know that any \(\mathtt{cx}\) gate commutes with a \(\mathtt{x}\) gate that acts on different qubits. The behavior of these gates is however different when they act on the same qubits.

Consider the following. Let \(\mathtt{x}_1\) and \(\mathtt{cx}_{12}\) act on each of the state \(\ket{00}\), \(\ket{01}\), \(\ket{10}\) and \(\ket{11}\) in both orders. Then, simple computations lead to 
\[ \begin{aligned} 
\mathtt{cx}_{12} \mathtt{x}_1 \ket{00} & = \mathtt{cx}_{12} \ket{10} = \ket{11}, \qquad \qquad \mathtt{x}_1 \mathtt{cx}_{12} \ket{00} = \mathtt{x}_1 \ket{00} = \ket{10},  \\
\mathtt{cx}_{12} \mathtt{x}_1 \ket{01} & = \mathtt{cx}_{12} \ket{11} = \ket{10}, \qquad \qquad  \mathtt{x}_1 \mathtt{cx}_{12} \ket{01} = \mathtt{x}_1 \ket{01} = \ket{11}, \\
\mathtt{cx}_{12} \mathtt{x}_1 \ket{10} & = \mathtt{cx}_{12} \ket{00} = \ket{00}, \qquad \qquad \mathtt{x}_1 \mathtt{cx}_{12} \ket{10} = \mathtt{x}_1 \ket{11} = \ket{01}, \\
\mathtt{cx}_{12} \mathtt{x}_1 \ket{11} & = \mathtt{cx}_{12} \ket{01} = \ket{01}, \qquad \qquad \mathtt{x}_1 \mathtt{cx}_{12} \ket{11} = \mathtt{x}_1 \ket{10} = \ket{00}. 
\end{aligned} \]
As anticipated, we have that \(\mathtt{x}_1\) and \(\mathtt{cx}_{12}\) do not commute. However, we can see that the results obtained differ exactly by a bitflip on the rightmost bit. Therefore, we can write
\[
\mathtt{cx}_{12} \mathtt{x}_1 \mathtt{x}_2 = \mathtt{x}_1 \mathtt{cx}_{12}. 
\]
We obtain similar result when applying \(\mathtt{x}_2\) or \(\mathtt{x}_1 \mathtt{x}_2\) after the \(\mathtt{cx}_{12}\): 
\[
\mathtt{x}_2 \mathtt{cx}_{12} = \mathtt{cx}_{12} \mathtt{x}_2, \qquad \qquad \mathtt{x}_1 \mathtt{x}_2 \mathtt{cx}_{12} = \mathtt{cx}_{12} \mathtt{x}_1.
\]
The circuit diagrams corresponding to these relations are given in Figure~\ref{fig: relations}. 

\begin{center}
\begin{figure}[ht]
\centering
    \begin{subfigure}{0.3\textwidth}
    \centering
    \begin{quantikz}[row sep = {0.7cm,between origins}, column sep = 0.25cm]
        \qw & \gate{\mathtt{x}} & \ctrl{1} & \qw \midstick[2,brackets=none]{=} & \ctrl{1} & \gate{\mathtt{x}} & \qw \\
        \qw & \gate{\mathtt{x}} & \targ{}  & \qw                                                 & \targ{}  & \qw                     & \qw
    \end{quantikz}
    \end{subfigure}
    \hspace{0.5cm}
    \begin{subfigure}{0.3\textwidth}
    \centering
    \begin{quantikz}[row sep = {0.7cm,between origins}, column sep = 0.25cm]
        \qw & \qw                     & \ctrl{1} & \qw \midstick[2,brackets=none]{=} & \ctrl{1} & \qw                      & \qw \\
        \qw & \gate{\mathtt{x}} & \targ{}  & \qw                                                 & \targ{}  & \gate{\mathtt{x}} & \qw
    \end{quantikz}
    \end{subfigure}
    \hspace{0.5cm}
    \begin{subfigure}{0.3\textwidth}
    \centering
    \begin{quantikz}[row sep = {0.7cm,between origins}, column sep = 0.25cm]
        \qw & \gate{\mathtt{x}} & \ctrl{1} & \qw \midstick[2,brackets=none]{=} & \ctrl{1} & \gate{\mathtt{x}}  & \qw \\
        \qw & \qw                     & \targ{}  & \qw                                                 & \targ{}  & \gate{\mathtt{x}} & \qw
    \end{quantikz}
    \end{subfigure}
    \caption{Relations between \(\mathtt{cx}\) and \(\mathtt{x}\) introduced in the new group.} 
    \label{fig: relations}
\end{figure} 
\end{center}

Interpreting these relations as rewriting rules, it is possible to move as far as one wishes the \(\mathtt{x}\) gates through \(\mathtt{cx}\) gates, with possibly a different number of \(\mathtt{x}\) gates at the end of the process. This eventually leads to the statement of Theorem~\ref{th: decomposition}.

A usual way for a group \(H\) to act on another group \(G\) is by \emph{conjugation}, i.e., via a group homomorphism \(\varphi_h\), defined for all element \(h \in H\) by \( \varphi_h(g) = hgh^{-1}\), for all \(g \in G\). Expressing the relations expressed above and in Figure~\ref{fig: relations} as conjugations yields
\[
\varphi_{jk} (\mathtt{x}_j) = \mathtt{x}_j \mathtt{x}_k, \qquad \qquad \varphi_{jk} (\mathtt{x}_k) = \mathtt{x}_k, \qquad \qquad \varphi_{jk}( \mathtt{x}_j \mathtt{x}_k) = \mathtt{x}_j, 
\]
with \(\varphi_{jk} = \varphi_{\mathtt{cx}_{jk}}\), remembering that \(\mathtt{cx}_{jk}^{-1} = \mathtt{cx}_{jk}\). Notice that \(\varphi_{jk}\) is a group morphism, illustrated for instance by \(\varphi_{jk}(\mathtt{x}_j \mathtt{x}_k) = \varphi_{jk}(\mathtt{x}_j) \varphi_{jk}(\mathtt{x}_k) = \mathtt{x}_1 \mathtt{x}_2 \mathtt{x}_2 = \mathtt{x}_1\). To be complete, we can add that \(\varphi_{kl}\) acts like the identity on \(\mathtt{x}\) gates acting on different qubits than \(k\) and \(l\). We now can give a presentation of the group \(LX_q\) generated by local permutation gates. 

\begin{proposition}\label{prop: presentation-lx}
A presentation of the group \(LX_q\) is given by the generators 
\[
\{ \mathtt{x}_j \mid 0 \leq j \leq q-1\} \cup \{ \mathtt{cx}_{kl} \mid k \neq l, 0 \leq k, l \leq q-1 \}
\]
and the relations given by the group \(X_q\), the group \(CX_q\) and the relations in Figure~\ref{fig: relations}: 
\begin{itemize}
\setlength\itemsep{0.5pt}
\item commutation: \( ( \mathtt{x}_j \mathtt{cx}_{kl} )^2 = \mathtt{I} \) where \(k \notin \{j, l\}\), and 
\item non-commutation: \( ( \mathtt{x}_j \mathtt{cx}_{jk} )^2 = \mathtt{x}_k \) where \(j \neq k\). 
\end{itemize}
\end{proposition}

\begin{proof}
We summarize the computation derived above. First, because \(LX_q\) is generated both by \(\mathtt{x}\) and \(\mathtt{cx}\) gates, then it satisfies the relations of \(X_q\) and \(CX_q\). For the additional ones expressed in Figure~\ref{fig: relations}, commutation relations arise whenever two gates are applied on different qubits, as expected. The last relations are derived exhaustively by trying all possible products of an individual \(\mathtt{cx}\) gate and an individual \(\mathtt{x}\) gate. There are four cases: 
\[
\mathtt{x}_j \mathtt{cx}_{jk}, \qquad \mathtt{x}_k \mathtt{cx}_{jk}, \qquad \mathtt{cx}_{jk} \mathtt{x}_j, \qquad \text{and} \qquad \mathtt{cx}_{jk} \mathtt{x}_k. 
\]
Studying the transformation induced by each of these circuits yields the identities: 
\[
\mathtt{x}_j \mathtt{cx}_{jk} = \mathtt{cx}_{jk} \mathtt{x}_j \mathtt{x}_k, \qquad \mathtt{x}_k \mathtt{cx}_{jk} = \mathtt{cx}_{jk} \mathtt{x}_k, \qquad \text{and} \qquad \mathtt{x}_j \mathtt{x}_k \mathtt{cx}_{jk} = \mathtt{cx}_{jk} \mathtt{x}_j. 
\]
The expression in the middle gives another commutation relation, which completes their enumeration. On the other ones, we multiply by \(\mathtt{cx}_{jk}\) on the right on all these expressions:
\[
\mathtt{x}_j = \mathtt{cx}_{jk} \mathtt{x}_j \mathtt{x}_k \mathtt{cx}_{jk} = \varphi_{jk}(\mathtt{x}_j \mathtt{x}_k), \qquad \text{and} \qquad \mathtt{x}_j \mathtt{x}_k = \mathtt{cx}_{jk} \mathtt{x}_j \mathtt{cx}_{jk} = \varphi_{jk}( \mathtt{x}_j ). 
\]
Because \(\varphi_{jk}( \mathtt{x}_k ) = \mathtt{x}_k\) and \(\varphi_{jk}\) is a group morphism, we can derive \(\varphi_{jk}(\mathtt{x}_j \mathtt{x}_k)\) from \(\varphi_{jk} ( \mathtt{x}_j )\): 
\[
\varphi_{jk} ( \mathtt{x}_j \mathtt{x}_k ) = \varphi_{jk} (\mathtt{x}_j) \varphi_{jk}(\mathtt{x}_k) = \mathtt{x}_j \mathtt{x}_k \mathtt{x}_k = \mathtt{x}_j, 
\]
therefore only \(\varphi_{jk}(\mathtt{x}_j) = \mathtt{cx}_{jk} \mathtt{x}_j \mathtt{cx}_{jk} = \mathtt{x}_j \mathtt{x}_k \) is necessary to complete the presentation. To put it under the standard form of a presentation, we rewrite it by multiplying on the left by \(\mathtt{x}_j\) on both sides, giving the non-commutation relations. 
\end{proof}

We now have a presentation of the group \(LX_q\), which is the group of all circuits on \(q\) qubits only composed of \(\mathtt{x}\) and \(\mathtt{cx}\) gates. Each element of \(LX_q\) represents such a circuit, up to the rewriting rules expressed as the relations of the presentation. 

For the time being, we still do not know how to efficiently expressed elements of \(LX_q\) as quantum circuits. So, to complete the study of the structure of the group \(LX_q\), we present a decomposition result for its elements \(U\) into two elements \(U = U_1 U_2\), with \(U_1\) written only using \(\mathtt{x}\) gates, and \(U_2\) only by \(\mathtt{cx}\) gates. This decomposition result follows from the \emph{semi-direct product} structure of \(LX_q\). 

\begin{definition}[{Semi-direct product~\cite[p.~177]{dummit2004abstract}}]
Let \(N\) and \(H\) be groups and let \(\varphi\) be a group morphism from \(H\) into \(\mathrm{Aut}(N)\)\footnote{As usual $\mathrm{Aut}(N)$ denotes the group of all possible automorphisms of a given group $N$.}. Then the (outer) \emph{semi-direct product} \(N \rtimes_\varphi H\) of \(N\) and \(H\) with respect to \(\varphi\) is defined as follows:
\begin{itemize}
\setlength\itemsep{0.5pt}
\item The underlying set is the Cartesian product \(N \times H\), 
\item The group operation is determined, for all \(n_1, n_2\) in \(N\) and \(h_1, h_2\) in \(H\), by
\[ \begin{aligned}
( N \rtimes_\varphi H) \times (N \rtimes_\varphi H) & \to N \rtimes_\varphi H \\
\big( (n_1, h_1), (n_2, h_2) \big) & \mapsto \big( n_1 \varphi_{h_1} \hspace{-2pt} (n_2), \, h_1h_2 \big).
\end{aligned} \]
\end{itemize}
When \(\varphi\) is clear from the context, we simply write \(N \rtimes H\). 
\end{definition}

The semi-direct product of two groups can also be computed using the presentations of the two factor groups \(N\) and \(H\) and a choice of group morphism \(\varphi: H \to \mathrm{Aut}(N)\). 

\begin{proposition}[{\cite[Corollary~1, p.140]{johnson1997presentations}}]
Let \(H = \pres{X}{R}\) and \(N = \pres{Y}{S}\) be groups, and \(\varphi: H \to \mathrm{Aut}(N)\) a group homomorphism. Then the semidirect product \( N \rtimes_\varphi H \) has the following presentation:
\[
N \rtimes H = \pres{X \cup Y}{R, S, x y x^{-1} = \varphi(x)(y), \forall x \in X, y \in Y}. 
\]
\end{proposition}

We can now derive the relationship among the three groups \(X_q\), \(CX_q\) and \(LX_q\). 

\begin{proposition}\label{prop: semi-direct-product}
The group of $\mathtt{x}$ and $\mathtt{cx}$ gates is isomorphic to the semi-direct product of the groups of $\mathtt{x}$ and $\mathtt{cx}$ gates taken alone, that is:  \(LX_q \simeq X_q \rtimes CX_q\). 
\end{proposition}

\begin{proof}
We use the same notation as before. We indeed have that \(LX_q\) is generated by the union of the generators of \(X_q\) and \(CX_q\). Moreover, the elements of \(LX_q\) satisfy both the relations of \(X_q\) and \(CX_q\). Now, we set 
\[ \begin{array}{rcrcl}
\varphi: CX_q & \to & \mathrm{Aut}(X_q) & & \\
\mathtt{cx}_{kl} & \mapsto & \varphi_{kl}: X_q & \to & X_q \\
& & w & \mapsto & \varphi_{kl}(w) = \mathtt{cx}_{kl} w \mathtt{cx}_{kl}~.
\end{array} \]
The automorphisms \(\varphi_{kl}\) already act by conjugacy, and the relations of \(LX_q\) contains all the relations that can be generated from it, which are the ones described in Proposition~\ref{prop: presentation-lx}. Then, the presentation of \(X_q \rtimes CX_q\) coincides with the one of \(LX_q\), and because presentations characterize groups up to isomorphism, we have that \(LX_q \simeq X_q \rtimes CX_q\). 
\end{proof}

It turns out that the group \(LX_q\) is isomorphic to a well-known group. Let \(V\) be a vector space. We call the group \(\mathrm{AGL}_n(V)\) defined by \(\mathrm{AGL}_n(V) = V \rtimes \mathrm{GL}_n(V)\) the \emph{affine group} of \(V\).

\begin{corollary}
The group of permutations generated by \(\mathtt{x}\) and \(\mathtt{cx}\) gates is isomorphic to the affine group of \(\mathbb{F}_2^q\), i.e., \(LX_q \simeq \mathrm{AGL}_q(\mathbb{F}_2)\). 
\end{corollary}

\begin{proof}
This follows from Proposition~\ref{prop: isomorphism-linear} and the definition of the affine group.
\end{proof}

\begin{remark}\label{rem: normal-subgroup}
Notice that the roles of \(X_q\) and \(CX_q\) here are \emph{asymmetric}. Indeed, elements of \(CX_q\) are mapped to automorphisms of \(X_q\) in a natural way, and this is possible because, for all \(\mathtt{cx}_{kl} \in CX_q\), we have that \(\varphi_{kl} (w) \in X_q\) for all \(w \in X_q\). We can extend this to the whole group, by noticing that \(w_1 w_2 w_1^{-1} \in X_q\) for all \(w_1, w_2 \in X_q\). Groups satisfying this properties are said to be \emph{normal} in their ambient group, and we write \(X_q \trianglelefteq LX_q\). 
\end{remark}

\medskip 

The next result provides the last tool necessary for the derivation of the decomposition. 

\begin{proposition}[{\cite[Theorem~12 and Proposition~8]{dummit2004abstract}}]\label{prop: equivalence-product}
Suppose \(G\) is a group with subgroups \(N\) and \(H\) such that \(N \trianglelefteq G\), and \(N \cap H = \mathrm{id}\). Let \(\varphi: H \to \mathrm{Aut}(N)\) be the automorphism defined by mapping \(h \in H\) to the automorphism \(\varphi_h: n \in N \mapsto hnh^{-1} \in N\). Then \(NH \simeq N \rtimes H\). Moreover, each element of \(NH\) can be written uniquely as a product \(nh\), for some \(n \in N\) and \(h \in H\).
\end{proposition}

The result above summarizes what it means to be a semi-direct product for two groups. Simply, we can take products of elements of each groups, and form an element of a larger group. However, characterizing the group by identifying all the relations satisfied by the elements is not trivial whenever some of the elements do not commute. We are now able to state our decomposition result. 

\begin{theorem}\label{th: decomposition}
	Every element \(U \in LX_q\) can be uniquely written as a product \(U_1 U_2\), with \(U_1 \in X_q\) and \(U_2 \in CX_q\). 
\end{theorem}

\begin{proof}
	By Proposition~\ref{prop: semi-direct-product}, we know that \(LX_q \simeq X_q \rtimes CX_q\), so we aim to apply Proposition~\ref{prop: equivalence-product}. By Remark~\ref{rem: normal-subgroup}, we have that \(X_q\) is a normal subgroup of \(LX_q\). The last hypothesis to check is that the intersection of \(X_q\) and \(CX_q\) is the identity. Consider any element \(w_X \in X_q\) which is not the identity, and apply it to the ground state of a \(q\)-qubit register, \(\ket{0}_q\). Necessarily, we have that \(w_X \ket{0}_q \neq \ket{0}_q\). Now, let \(w_{CX}\) be an element of \(CX_q\) not being the identity. Necessarily, we have that \(w_{CX} \ket{0}_q = \ket{0}_q\). Then, \(w_X \neq w_{CX}\), and the only common element between \(X_q\) and \(CX_q\) is the identity. We can now apply Proposition~\ref{prop: equivalence-product}, which concludes the proof. 
\end{proof}

Thanks to Theorem~\ref{th: decomposition}, it suffices to know how to design a circuit \(\mathscr{C}_X\) that spans the entirety of \(X_q\) and another circuit \(\mathscr{C}_{CX}\) that spans the entirety of \(CX_q\) to be able to construct a quantum algorithm spanning every possible circuit formed by \(\mathtt{x}\) and \(\mathtt{cx}\) gates, up to the rewriting rules. The design of these circuits is the topic of the next section.

\section{Quantum circuits}\label{sec:circuits}

\subsection{Circuits for $X_q$}

In this section, we translate the group-theoretic results from the previous session into concrete tools to build circuits exploring groups. A straightforward example is the circuit \(\mathscr{C}_X(\theta)\) exploring the group \(X_q \simeq \mathbb{F}_2^q\). It simply amounts to a layer of \(\mathtt{rx}(\theta_j)\) gates, with the parameter \(\theta_j\) playing the role of a switch: when it is evaluated at \(0\), the gate is the identity, and when it is evaluated at \(\pi\), then the acts like \(\mathtt{x}\), up to an unimportant global phase factor. The strength of the variational quantum circuit \(\mathscr{C}_X(\theta)\) is that it allows for more value of \(\theta_j\) than just \(0\) and \(\pi\). Actually, any real number is a valid parameter. Hence, the circuit explores the group \(X_q\) while travelling through some extra space giving by this relaxation, therefore being in a superposition of permutations.  

\subsection{Circuits for $CX_q$}

The picture gets blurrier when we want to explore the group \(CX_q\). We need to leverage the results of the previous section to divide this problem into easier subproblems, i.e., smaller groups for which we know how to design variational quantum circuits exploring them. 

We recall that it has been shown by Bataille~\cite{bataille2022quantum} that \(CX_q\) is isomorphic to \(\mathrm{GL}_q(\mathbb{F}_2)\). In order to span the latter group in a compact way with a parameterized circuit, we use the Bruhat decomposition of \(\mathrm{GL}_q(\mathbb{F}_2)\) (see Theorem~\ref{th: bruhat-decomposition}). Consider the following. 

Given a group \(G\) with subgroups \(B\) and \(N\), the data defining a \(BN\)-pair~\cite{tits1974buildings} consists formally of a quadruple \((G, B, N, S)\) subject to the following requirements. First, \(G\) is generated by its subgroups \(B\) and \(N\). Second, \(B \cap N\) is a normal subgroup of \(N\). Third, the group \(W \coloneqq N / (B \cap N)\) is generated by a set \(S\) of involutions and, finally, for all \(s \in S\) and \(w \in W\), we have \(s B w \subseteq B w B \cup B s w B\), as well as \(s B s \neq B\). 

In the case \(G = \mathrm{GL}_q(\mathbb{F}_2)\), the subgroup \(B\) is the \emph{Borel subgroup} formed by the upper triangular matrices, and \(N\) is the subgroup having a unique non-zero element per row and per column~\cite{bourbaki1968groupes}. Over \(\mathbb{F}_2\), \(N\) is only composed of permutation matrices. Then, the subgroup \(B \cap N = \{\mathtt{I}\}\) is trivial, because the only permutation matrix being upper triangular is the identity. Therefore, the \emph{Weyl group} \( W = N / (B \cap N) \) of the \(BN\)-pair is simply the symmetric group \(S_q\). 

Recall now that the Bruhat decomposition theorem states that any group \(G\) satisfying the conditions described above can be written as \(G = BWB\). More specifically, each element of \(\mathrm{GL}_q(\mathbb{F}_2)\), and therefore of \(CX_q\), can be uniquely written as the product of a permutation matrix, multiplied on the left and the right by upper triangular matrices. Fortunately, both \(W\) and \(B\) can be easily embedded into quantum circuits. 

\subsection{The Borel subgroup $B$}

The Borel subgroup \(B\) is the subgroup of \(\mathrm{GL}_q(\mathbb{F}_2)\) composed of upper triangular matrices. Because they are invertible, and defined over \(\mathbb{F}_2\), their diagonal only contains the entry \(1\). Hence, their coefficients can only differ strictly above the diagonal. The matrices are of size \(q \times q\), so there are \((q-1) + (q-2) + \ldots + 1 = \binom{q}{2}\) entries that vary. Because each of them can only take two values, there are \(2^{\binom{q}{2}}\) elements in the Borel subgroup \(B\). Hence, the minimal number of gates we need to span \(B\) is \(\binom{q}{2}\). 

Because we are on the field \(\mathbb{F}_2\), the group \(\mathrm{GL}_q(\mathbb{F}_2)\) is isomorphic to \(\mathrm{SL}_q(\mathbb{F}_2)\), the group of invertible matrices with determinant \(1\). Therefore, we can use the result presented in~\cite[Theorem~4.6]{artin2016geometric} to say that a set of generators is given by the transvections 
\[
\{T_{(jk)} \mid 1 \leq j, k \leq q\}, \qquad \text{with} \qquad T_{(jk)} = I + E_{(jk)}, 
\]
where \(E_{(jk)}\) is the matrix with a \(1\) at entry \((j, k)\) and \(0\) everywhere else. Bataille~\cite{bataille2022quantum} has shown that the transvections are in bijection with the \(\mathtt{cx}\) gates. More precisely, the transvection \(T_{(jk)}\) corresponds to the \(\mathtt{cx}\) gate controlled by qubit \(k\) acting on qubit \(j\). This construction is used to build the isomorphism between \(\mathrm{GL}_q(\mathbb{F}_2)\) and \(CX_q\). 
 
As we will demonstrate shortly, the Borel subgroup \(B\) consisting of upper-triangular invertible matrices is generated by upper-triangular transvections, i.e., by the set 
\[
\left\{ T_{(jk)} \mid 1 \leq j < k \leq q \right\}.
\]
Then, the group \(B\) is generated by the \(\mathtt{cx}\) gates where the index of the control qubit is smaller than the one of the target qubit. However, to build a circuit able to span all elements of \(B\), we need to find a \emph{universal element} of the group, an element \(g_\star \in B\) which can be written as a word \(w_\star\) of transvections with the following property: for any element \(g \in B\) written as a word \(w\) of transvections, \(w\) is a subword of \(w_\star\). This is a direct generalization of the concept of longest element of the Bruhat order in Coxeter groups.

\begin{definition}
We say that a word \(w = s_{j_1} \ldots s_{j_p}\) is a subword of another word \(v = s_{k_1} \ldots s_{k_m}\) if there exists an order-preserving injection from the sequence \((j_1, \ldots, j_p)\) into the sequence \((k_1, \ldots, k_m)\). In other words, \(w\) can be written by taking arbitrary letters from \(v\) as long as the order between them does not change. For example, \(ade\) is a subword of \(abcde\), while \(acb\) or \(abf\) are not.
\end{definition}

Let \((Q, \prec)\) be the ordered set with \(Q = \{(j, k) \mid 1 \leq j < k \leq q\}\) being composed of all the ordered pairs of increasing indices equipped with the lexicographic order \(\prec\) given by 
\[
(j_1, k_1) \prec (j_2, k_2) \qquad \text{ if and only if } \qquad \left[ j_1 < j_2 \right] \text{ or } \left[ j_1 = j_2 \text{ and } k_1 < k_2 \right]. 
\]
The order \(\prec\) is a total order on \(Q\). We denote by \(w_\star\) the word of transvections defined as the product of all transvections multiplied on the left in the order of \(\prec\), i.e., 
\[
w_\star \coloneqq T_{(q-1 \hspace{2pt} q)} T_{(q-2 \hspace{2pt} q-1)} T_{(q-2 \hspace{2pt} q)} \ldots T_{(jq)} \ldots T_{(j \hspace{2pt} j-1)} T_{(j-1 \hspace{2pt} q)} \ldots T_{(23)} T_{(1q)} \ldots T_{(13)} T_{(12)}
.
\]

\begin{theorem}\label{th: transvection-generation}
    Any element of \(B\) can be written as a subword of \(w_\star\). 
\end{theorem}

\begin{proof}
Let \(A\) be an upper triangular matrix of \(B\). We show that we can obtain \(A\) from the identity matrix by multiplying it on the left by a subword of \(w_\star\). We construct it entry by entry, in the order given by \(\prec\). Let \((Q_A, \prec)\) be the ordered set of indices defined by \( Q_A = \{(j, k) \in Q \mid A_{jk} = 1\} \). We denote by \(w_A\) the subword of \(w_\star\) obtained by only keeping the symbols \(\{T_{(jk)} \mid (j, k) \in Q_A\}\). Let \(M \in B\) be a matrix and \((p, r)\) be a pair of indices such that, for all \((p, r) \prec (j, k) \in Q\) we have \(M_{jk} = 0\), as well as \(M_{pr} = 0\). We have 
\[
T_{(pr)} M = \left( I + E_{(pr)} \right) M = M + E_{(pr)} M.
\]
Because \(E_{(pr)}\) has a unique nonzero coefficient at entry \((p, r)\), the product \(E_{(pr)} M\) only consists of the \(r\)th row of \(M\) sitting on the \(p\)th row. But because \(M_{jk} = 0\) for all \((j, k) \succ (p, r)\), the \(r\)th row of \(M\) is simply \(e_r^\top\), the transpose of the \(r\)th element of the canonical basis of \(\mathbb{F}_2^q\). Then, \(T_{(pr)} M\) is equal to \(M\) to which we add \(1\) at the entry \((p, r)\). Because we only apply transvections following the lexicographic order \(\prec\), the product \(w_A\) adds entries \(1\) one by one to the identity when the corresponding entry in \(A\) is itself \(1\), so it indeed corresponds to the matrix \(A\). 
\end{proof}

\begin{example} We illustrate this result with a specific example. Let \(A\) be an upper-triangular matrix defined by 
\[
A = \begin{psmallmatrix} 
1 & 0 & 1 & 0 \\ 
0 & 1 & 1 & 1 \\
0 & 0 & 1 & 0 \\
0 & 0 & 0 & 1
\end{psmallmatrix}.
\]
Then, \(A\) can be written as \( A = T_{24} T_{23} T_{13} \). Indeed, we have
\[
T_{24} \begin{psmallmatrix}
1 & 0 & 0 & 0 \\
0 & 1 & 1 & 0 \\
0 & 0 & 1 & 0 \\
0 & 0 & 0 & 1
\end{psmallmatrix} \begin{psmallmatrix} 
1 & 0 & 1 & 0 \\
0 & 1 & 0 & 0 \\
0 & 0 & 1 & 0 \\
0 & 0 & 0 & 1
\end{psmallmatrix} = \begin{psmallmatrix} 
1 & 0 & 0 & 0 \\
0 & 1 & 0 & 1 \\
0 & 0 & 1 & 0 \\
0 & 0 & 0 & 1
\end{psmallmatrix} \begin{psmallmatrix} 
1 & 0 & 1 & 0 \\
0 & 1 & 1 & 0 \\
0 & 0 & 1 & 0 \\
0 & 0 & 0 & 1
\end{psmallmatrix} = \begin{psmallmatrix} 
1 & 0 & 1 & 0 \\
0 & 1 & 1 & 1 \\
0 & 0 & 1 & 0 \\
0 & 0 & 0 & 1
\end{psmallmatrix} = A.  
\]
\end{example}

The next step is to apply Theorem~\ref{th: transvection-generation} to actually build a quantum circuit $\mathscr{B}(\theta)$. Using the correspondence between \(\mathtt{cx}\) gates and transvections, we are able to design a variational circuit exploring the Borel subgroup \(B\) of \(CX_q\). In these circuits, we use a custom gate, displayed as a parameterized box with a control wire. This gate acts as the identity whenever the parameter evaluates to \(0\), and as a \(\mathtt{cx}\) gate when the parameter is \(\pi\), although it is defined for all real-valued parameters. The construction is discussed in Appendix~\ref{ap.2}. 

\begin{example}\label{ex: borel}
Let us illustrate this first construction on \(6\) qubits. In this case, the Borel group \(B_6\) is the set of upper-triangular matrices of size \(6 \times 6\) equipped with the usual matrix product. Applying Theorem~\ref{th: transvection-generation}, we have that any upper-triangular matrix can be written as a subword of
\[
w_\star^{(6)} = \hspace{4pt} T_{(56)} \hspace{5pt} T_{(46)} T_{(45)} \hspace{5pt} T_{(36)} T_{(35)} T_{(34)} \hspace{5pt} T_{(26)} T_{(25)} T_{(24)} T_{(23)} \hspace{5pt} T_{(16)} T_{(15)} T_{(14)} T_{(13)} T_{(12)}. 
\]
Consider now the implementation of an upper-triangular matrix \(M \in B_6\) as a quantum circuit. We first look at the entries of our matrix from the bottom left to the bottom right, line per line. We see that we have a transvection matrix for each of the entries. If the entry \(M_{jk}\) is \(0\), we remove the corresponding transvection \(T_{jk}\) from \(w_\star^{(6)}\), but if it is \(1\), we keep it. The word that we then obtain is equal to the matrix \(M\). To translate it into a quantum circuit, we use the isomorphism of Bataille~\cite{bataille2022quantum}. We read the word representing \(M\) from right to left, and for each transvection \(T_{(jk)}\), we add a \(\mathtt{cx}_{kj}\) gate to our circuit. Hence, any upper-triangular matrix is implement as a subcircuit of the circuit spanning the Borel subgroup \(B_6\) depicted in Figure~\ref{fig: first-borel}. 

\begin{figure}[H]
    \begin{center}
	\includegraphics[width=0.95\textwidth]{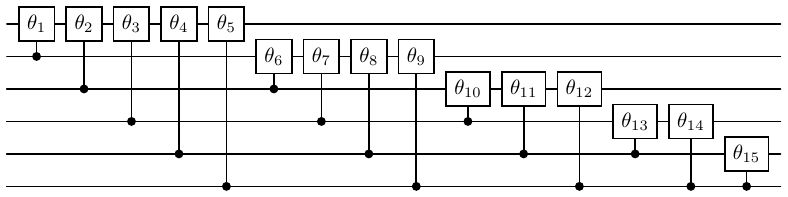}
    \end{center}
    \caption{Variational circuit $\mathscr{B}(\theta)$ spanning the Borel subgroup \(B_6\).}
    \label{fig: first-borel}
\end{figure}
The parameters \(\theta\) in the circuits control the application of the corresponding gate in the circuit: putting \(\theta = 0\) acts as if we remove the gate, while \(\theta=\pi\) implements it fully. The cases when \(\theta\) is neither \(0\) nor \(\pi\) is treated in the next section. Notice how the order is flipped: this is due to the fact that functions from formulas are applied from the right to the left, but the opposite in circuit diagrams.
\end{example}

The number of parameters and gates required in this circuit is in \(O(q^2)\). However, some of these gates can be applied in parallel. For instance, the gate depending on \(\theta_6\) can be applied as early as the one depending on \(\theta_3\), and the one depending on \(\theta_{10}\) as early as the one depending on \(\theta_8\), which is itself applied with the gate depending on \(\theta_5\). 

We can see the circuit as composed of several blocks, the first one being composed of all the \(\mathtt{cx}\) gates controlled by the first qubit, the second block being composed of the \(\mathtt{cx}\) gates controlled by the second qubit, and so on. We therefore have \(q-1\) blocks. The gates from the second block can all be executed in parallel of the one of the first block, except for the last one, the one depending on \(\theta_9\). The same phenomenon happens with each block, where all but the last gate can be applied with the ones of the previous blocks. 

Hence, even if the size of the circuit is quadratic in the number of qubits, its depth is simply linear, as it is given by \( (q-1) + (q-2) \sim 2q \in O(q)\). Indeed, the first block is of depth \(q-1\), and each following bloc, \(q-2\) of them, only adds a depth of \(1\). 

\subsection{The Weyl subgroup $W$}

As we saw earlier, the Weyl subgroup \(W\) of the \(BN\)-pair is isomorphic to the symmetric group \(S_q\), which can be seen as the group of permutations of the \(q\) qubits (see also Example~\ref{example.1}). A presentation of the group \(W\) is therefore given by 
\[
S_q = \pres{\sigma_1, \ldots, \sigma_{q-1}\, }{\, \sigma_k^2 = 1, \; \left( \sigma_k \sigma_{k+1} \right)^3 = 1, \; \left( \sigma_k \sigma_j \right)^2 = 1 \text{ if } \lvert k - j \rvert > 1}, 
\]
with the \(\sigma_1, \ldots, \sigma_{q-1}\) being the adjacent transpositions defined by \(\sigma_k = (k \hspace{5pt} k+1)\), written using the cycle notation. Like the upper triangular matrices being written as words of transvections, permutations can be written as words of adjacent transpositions. 

To implement these adjacent transpositions of qubits, we use the well-known decomposition of the \(\mathtt{swap}\) gates using three \(\mathtt{cx}\), the first and the third one pointing in a different direction than the second one (see Figure~\ref{fig: pswap}). The first and third gates are not parameterized, but the second one is. By doing so, the whole \(\mathtt{swap}\) gate collapses if the middle \(\mathtt{cx}\) gate is set to be the identity, with \(\phi = 0\). However, when \(\psi = \pi\), the parameterized \(\mathtt{cx}\) gate becomes a standard \(\mathtt{cx}\) gate, and the three gates together are indeed a \(\mathtt{swap}\) gate. 

\begin{figure}[H]
    \begin{center}
    \begin{quantikz}[align equals at=1.5, row sep = {0.6cm,between origins}, column sep = 0.3cm]
        \ghost{\targ{}} & \permute{2, 1}\gategroup[2, steps = 1, style = {transparent}, label style = {label position = below, anchor = south, yshift = 0.4cm, xshift = 0.35cm}]{$\phi$} & \qw \\
        \ghost{\phi} & & \qw 
    \end{quantikz} = 
    \begin{tikzcd}[align equals at=1.5, row sep = {0.6cm,between origins}, column sep = 0.3cm]
        \ghost{\targX{}} & \targ{} & \ctrl{1} & \targ{} & \qw \\
        \ghost{\targX{}} & \ctrl{-1} & \gate{\phi} & \ctrl{-1} & \qw 
    \end{tikzcd}
    \end{center}
    \vspace{-10pt}
    \caption{Construction of a parameterized \(\mathtt{swap}\) gate.}
    \label{fig: pswap}
\end{figure}

We now gather all these adjacent transpositions in a harmonious way in order to generate any required permutation of qubits, which compose the Weyl subgroup \(W\). 

The partial order on word representations of permutations obtained by subword relations corresponds to the Bruhat order on Coxeter groups, also known as the weak order of permutations in the case of the symmetric group. Each Bruhat order has a unique maximal element. Then, to span the whole group \(S_q\), we implement a word representation of this longest element of the Bruhat order as a quantum circuit. 
 
In particular, the following characterization of the Bruhat order on the symmetric group allows us to identify this longest element. For each \(\eta \in S_q\) and each \(j, k \in [q]\), define
\[
\eta[j, k] \coloneqq \lvert \{ l \in [j] \mid \eta(l) \geq k\} \rvert.
\]
This function counts, for each index \(j \in [q]\), the number of smaller indices sent after index \(k\) through the permutation \(\eta\). We then use the following result. 

\begin{theorem}[Björner and Brenti~\cite{bjorner2005combinatorics}]\label{th: bruhat-order}
    Let \(\eta, \rho \in S_q\) be two permutations. Then, \(\eta\) is smaller than \(\rho\) in the Bruhat order if and only if \(\eta[j, k] \leq \rho[j, k]\) for all \(j, k \in [q]\).
\end{theorem}

Naturally, the permutation \(\eta_\circ \in S_q\) reversing the order of the elements of its input is the unique permutation reaching the maximal value \(\eta_\circ[j, k]\) on each pair \(j, k \in [q]\). Then, by Theorem~\ref{th: bruhat-order}, it is the maximal element of the Bruhat order. A maximal element for the Bruhat order is necessarily unique, and in the case of \(S_q\), its length, as a word of adjacent transpositions, is \(\binom{q}{2}\)~\cite{bjorner2005combinatorics}. A well-known word representation for the permutation \(\eta_\circ\) is 
\[
w_\circ \coloneqq \hspace{4pt} \sigma_1 \hspace{5pt} \sigma_2 \sigma_1 \hspace{5pt} \sigma_3 \sigma_2 \sigma_1 \hspace{5pt} \ldots \hspace{5pt} \sigma_{q-1} \sigma_{q-2} \ldots \sigma_2 \sigma_1.
\]
It is obtain by first multiplying all the adjacent transpositions in the order of the indices, then multiplying on the left the product of all the adjacent transpositions in the order of the indices except the last one \(\sigma_{q-1}\), and we repeat this recursively, dropping the last transposition at each step. The \(q-1\) first transpositions applied, i.e., the ones on the right of \(w_\circ\), serve to put the first element of the input at the last spot. Then, the \(q-2\) next ones are used to put the second element of the input at the second-to-last spot, and so on. Then, \(w_\circ\) is a word representing the reversing permutation \(\eta_\circ\). Moreover, the length of \(w_\circ\) is given by \(\binom{q}{2}\), and is indeed one of the reduced word representation of \(\eta_\circ\).  

\newcommand{\pswap}[1]{\permute{2, 1}\gategroup[2, steps = 1, style = {transparent}, label style = {label position = below, anchor = south, yshift = 0.115cm, xshift = 0.38cm}]{#1}}

\begin{example}\label{ex: weyl}
Let us continue Example~\ref{ex: borel}, with \(6\) qubits. In this case, the elements of the Weyl group \(W_6\) are the permutations of \(6\) symbols, here the \(6\) qubits. Then, the maximal element of the Bruhat order of \(W_6\) is given by
\[
w_\circ^{(6)} = \hspace{4pt} \sigma_1 \hspace{5pt} \sigma_2 \sigma_1 \hspace{5pt} \sigma_3 \sigma_2 \sigma_1 \hspace{5pt} \sigma_4 \sigma_3 \sigma_2 \sigma_1 \hspace{5pt} \sigma_5 \sigma_4 \sigma_3 \sigma_2 \sigma_1.
\] 
The quantum circuit to span the Weyl subgroup \(W_6\) is depicted in Figure~\ref{fig: first-weyl}. 

\begin{figure}[H]
    \begin{center}
    \includegraphics[width=0.9\textwidth]{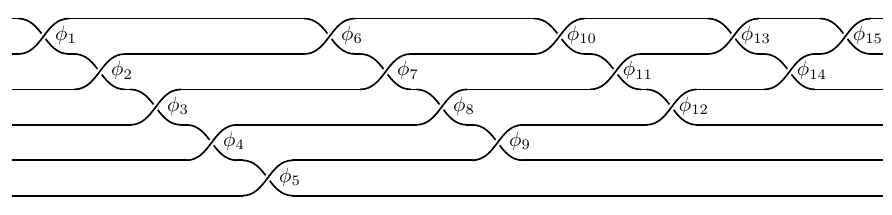}
    \end{center}
    \vspace{-10pt}
    \caption{Variational circuit $\mathscr{W}(\phi)$ spanning the Weyl subgroup \(W\).}
    \label{fig: first-weyl}
\end{figure}
Again, the parameters \(\phi\) enable the user to control if the gate is active or not, allowing the access to any subword.
\end{example}

The number of parameters of this circuit is \(\binom{q}{2} \in O(q^2)\), and the number of controlled gates is \(3 \binom{q}{2} \in O(q^2)\). However, as before, the depth is significantly shorter. Indeed, a lot of gates can be applied in parallel, in the same way than for the previous circuit. As before, there are \(q-1\) blocks, the first one increases the depth of the circuit by \(3(q-1)\), and the following ones increase the depth by \(3\). Hence, the depth is \(3(q-1) + 3(q-2) \in O(q)\). 

\subsection{Exploring the whole group generated by \texttt{cx} gates and $LX_q$}

To span the whole group \(CX_q\), thanks to the Bruhat decomposition, we simply need to concatenate the different circuits spanning \(B\) and \(W\). Therefore, the total number of controlled gates used to span \(CX_q\) is \(\binom{q}{2} + 3\binom{q}{2} + \binom{q}{2} = 5\binom{q}{2} \in O(q^2)\), while the number of parameters used is \(3\binom{q}{2} \in O(q^2)\). The total depth is less than the sum of the depths, because part of the gates of the circuit exploring \(W\) can be applied at the same time as some of the first occurrence of the circuit spanning \(B\). 

\renewcommand{\pswap}[1]{\permute{2, 1}\gategroup[2, steps = 1, style = {transparent}, label style = {label position = below, anchor = south, yshift = 0.41cm, xshift = 0.38cm}]{#1}}

\begin{example}\label{ex: bruhat}

Let us consider \(4\) qubits. According to Theorem~\ref{th: bruhat-decomposition}, the Bruhat decomposition of \(CX_4\) is given by \(B_4 W_4 B_4\). The circuit spanning \(CX_4\) is depicted in Figure~\ref{fig: first-bruhat}.

\begin{figure}[H]
    \begin{center}
    \begin{adjustbox}{width=\textwidth}
    \begin{tikzcd}[row sep = {0.6cm,between origins}, column sep = 0.2cm]
        \qw & \gate{\theta_1}\gategroup[4, steps=6, style={dashed, rounded corners, inner sep=-1pt}]{Circuit spanning \(B_4\)} & \gate{\theta_2} & \gate{\theta_3} & \qw & \qw & \qw & \qw & \pswap{$\phi_1$}\gategroup[4, steps=6, style={dashed, rounded corners, inner sep=-1pt}]{Circuit spanning \(W_4\)} & \qw & \pswap{$\phi_4$} & \qw & \pswap{$\phi_6$} & \qw & \qw & \gate{\theta_7}\gategroup[4, steps=6, style={dashed, rounded corners, inner sep=-1pt}]{Circuit spanning \(B_4\)} & \gate{\theta_8} & \gate{\theta_9} & \qw & \qw & \qw & \qw \\
        \qw & \ctrl{-1} & \qw & \qw & \gate{\theta_4} & \gate{\theta_5} & \qw & \qw & \qw & \pswap{$\phi_2$} & & \pswap{$\phi_5$} & \qw & \qw & \qw & \ctrl{-1} & \qw & \qw & \gate{\theta_{10}} & \gate{\theta_{11}} & \qw & \qw \\
        \qw & \qw & \ctrl{-2} & \qw & \ctrl{-1} & \qw & \gate{\theta_6} & \qw & \qw & \qw & \permute{2, 1}\gategroup[2, steps = 1, style = {transparent}, label style = {label position = below, anchor = south, yshift = 0.18cm, xshift = 0.38cm}]{$\phi_3$} & & \qw & \qw & \qw & \qw & \ctrl{-2} & \qw & \ctrl{-1} & \qw & \gate{\theta_{12}} & \qw \\
        \qw & \qw & \qw & \ctrl{-3} & \qw & \ctrl{-2} & \ctrl{-1} & \qw & \qw & \qw & \qw & \qw & \qw & \qw & \qw & \qw & \qw & \ctrl{-3} & \qw & \ctrl{-2} & \ctrl{-1} & \qw
    \end{tikzcd}
    \end{adjustbox}
    \end{center}
    \caption{Variational circuit spanning \(CX_4\).}
    \label{fig: first-bruhat}
\end{figure}
Notice here that the parametrization of the two Borel blocks are not parameterized by the same values: the two blocks can encode different upper-triangular matrices.
\end{example}

The final depth of the circuit is given by 
\[
q + 3(q-1) + 3(q-2) + (q-1) + (q-2) = 9q - 12 \in O(q), 
\]
hence is linear in the number of qubits. 

Finally, to explore the whole group \(LX_q\), it suffices to add a layer of \(\mathtt{rx}\) gate upstream of the circuit, which increases the size of the circuit by \(q\), and its depth by \(1\). 

\subsection{Topology constraints on near-term devices}

In the circuits presented above, we assume an all-to-all qubit connectivity. However, on some hardware, and especially for near-term processors, this assumption can be too demanding. For the circuit spanning the Weyl group \(W_q\), there should be no issue during the physical implementation in terms of connectivity: we are only using \(2\)-qubits gates acting on neighboring qubits. For the circuit spanning the Borel group \(B_q\), this is no longer true, because the (\(q\text{-}1\))-th gate has the longest possible range. To solve this issue, we use the method developed by B{\"a}umer and W{\"o}rner~\cite{baumer2025measurement} to implement long-range \(\mathtt{cx}\) gates. If we have access to ancilla qubits, we can direct use their implementation, which works in constant depth, hence it does not change the overall scaling of the depth of our circuits. For an ancilla-free implementation, we resort to the rewriting rules in Figure~2 and Figure~3 of~\cite{baumer2025measurement}, as follows.

A long-range \(\mathtt{cx}\) gate can be decomposed as two multi-target \(\mathtt{cx}\) gates, which can themselves be decomposed as many \(\mathtt{cx}\) acting on neighboring qubits as follows: 
\begin{figure}[h]
\begin{center}
    \begin{tikzcd}[row sep = {0.5cm,between origins}, column sep = 0.25cm]
        \qw & \targ{} & \qw \midstick[5,brackets=none]{=} & \targ{} & \qw      & \qw \midstick[5,brackets=none]{=} & \targ{}\gategroup[5, steps=7, style={dashed, rounded corners, inner sep=-1pt}]{First multi-target \(\mathtt{cx}\)}      & \qw & \qw & \qw & \qw & \qw & \targ{}    & \qw & \qw\gategroup[5, steps=5, style={dashed, rounded corners, inner sep=-1pt}]{Second one} & \qw & \qw & \qw & \qw & \\
        \qw & \qw      & \qw                                                 & \targ{} & \targ{} & \qw                                                  & \ctrl{-1} & \targ{}   & \qw       & \qw      & \qw       & \targ{}   & \ctrl{-1} & \qw & \targ{}   & \qw        & \qw      & \qw      & \targ{}  & \\ 
        \qw & \qw      & \qw                                                 & \targ{} & \targ{} & \qw                                                  & \qw      & \ctrl{-1}  & \targ{}   & \qw      & \targ{}   & \ctrl{-1} & \qw       & \qw & \ctrl{-1} & \targ{}    & \qw      & \targ{}  & \ctrl{-1} & \\
        \qw & \qw      & \qw                                                 & \targ{} & \targ{} & \qw                                                  & \qw      & \qw        & \ctrl{-1} & \targ{}  & \ctrl{-1} & \qw       & \qw       & \qw & \qw       & \ctrl{-1}  & \targ{}  & \ctrl{-1} & \qw      & \\
        \qw & \ctrl{-4} & \qw                                                & \ctrl{-4} & \ctrl{-3} & \qw                                               & \qw      & \qw       & \qw       & \ctrl{-1} & \qw       & \qw       & \qw       & \qw & \qw      & \qw        & \ctrl{-1} & \qw      & \qw      &  
    \end{tikzcd}
\end{center}
\caption{Decomposition of a long-range \(\mathtt{cx}\) gate using \(\mathtt{cx}\) acting on neighboring qubits.}
\label{fig: neighbor}
\end{figure}

Observe that, if we remove the two topmost \(\mathtt{cx}\) gates, all the remaining ones cancel, and we get the identity. So, we apply the parametrization on these \(\mathtt{cx}\) gates. It might not be exactly the same unitary gate as the long-range \(\mathtt{cx}\) for arbitrary angles, but it is not important, as long as the evaluation at \(\theta = 0\) or \(\theta = \pi\) are the same, meaning that we span the same permutations.

In terms of complexity, we get the following. When the index of target qubit is \(p\) less than the index of the control qubit, we need \(2(1 + 2(p-1)) - 2 = 4(p-1)\) neighboring \(\mathtt{cx}\) gates if \(p > 1\), and \(1\) if \(p=1\). So, in a circuit spanning the Borel group \(B_q\) on \(q > 3\) qubits, the number of long-range \(\mathtt{cx}\) gates targeting qubit \(j < q-2\) is
\[
1 + \sum_{p = 2}^{q - j} 4(p-1) = 1 + 4 \sum_{p' = 1}^{q-j-1} p' = 1 + 4 \binom{q-j}{2} \in O(q^2). 
\]
There are a linear number of these long-range \(\mathtt{cx}\), so the size and the depth of the circuit spanning the Borel group scale in \(O(q^3)\), giving an overall circuit of cubic size and depth.

\subsection{Counting the number of spanned permutations}

In this section, we have translated the group-theoretical results about the structures of the group \(LX_q\) of permutations generated by local permutations on \(q\) into tools for circuit synthesis. The meta-circuit exploring \(LX_q\) is of the form \(XBWB\), where \(X\) is the group of \(\mathtt{x}\) gates, \(B\) is the Borel subgroup of upper-triangular invertible matrices of \(\mathrm{GL}_q(\mathbb{F}_2)\), and \(W\) is the Weyl subgroup of \(\mathrm{GL}_q(\mathbb{F}_2)\), isomorphic to the group of permutations of \(q\) qubits. 

These circuits, using a quadratic number of parameters, are of linear depth with respect to the number of qubits. Thanks to the extensive characterization of the structure of \(LX_q\), we are able to compute the number of permutations spanned by the circuit. 

\begin{theorem}\label{th:number-of-permutations}
The maximum number of unique permutations $p$ obtainable by circuits containing \(\mathtt{x}\) and $\mathtt{cx}$ gates is  
\[
p = 2^{\frac{q(q+1)}{2}} \prod_{k = 1}^q \left( 2^k - 1 \right) \in O\left( 2^{O(q^2)} \right). 
\]
\end{theorem}

\begin{proof}
Given Theorem~\ref{th: decomposition}, we can count permutations by multiplying the ones coming from $X_q$ and the ones from $CX_q$. For the latter, we use the result~\cite[Corollary~7]{bataille2022quantum} about the number of elements of \(CX_q \simeq \mathrm{GL}_q(\mathbb{F}_2)\), which we then need to multiply by the number of elements of $X_q$ ($2^q$) to compute the number of elements of \(LX_q\).
\end{proof}

Evaluations of this formula for small numbers of qubits are given in Table~\ref{table: span}, along with the number of classical parameters $\ell = q + 3\binom{q}{2}$, and the depth $d = 1 + 9q - 12$. 

\begin{table}[h]
\caption{Number \(p\) of spanned permutations with \(q\) qubits, using \(\ell\) classical parameters, in $d$ depth. In comparison the size of the permutation set of $\Pi_n$ with $n = 2^q$, which we encode, is $n!$. }
\label{table: span}
\centering

\begin{tabular}{cccccccc}
    \toprule
    \(q\) \qquad & 2 & 3 & 4 & 5 & 6 & 7 \\
    \midrule
    \(\ell\) \qquad & 5 & 12 & 22 & 35 & 51 & 70 \\
    \(d\) \qquad & 7 & 16 & 25 & 34 & 43 & 52 \\
    \(p\) \qquad & 24 & 1,344 & 322,560 & \(\sim 3.2 \times 10^8\) & \(\sim 1.3 \times 10^{12}\) & \(\sim 2.1 \times 10^{16}\) \\
    \midrule 
    $\lvert \Pi_n \rvert$ \qquad & 24 & 40,320 & \(\sim 2.1 \times 10^{13}\) & $\sim 2.6 \times 10^{35}$ & $\sim1.2 \times 10^{89}$ & $\sim 3.8 \times 10^{215}$ \\ 
    \bottomrule
\end{tabular}
\end{table}

The number of spanned permutations, i.e., all permutations obtained from one- and two-qubit gates, is rather small compared to the total number of permutations (with the exception of $q=2$ for which the two numbers match). We see in the next section how to make use of ancilla qubits to both increase the span and to avoid classical simulability. Furthermore, we will see in the numerical simulation section, how, despite the small span, we obtain reasonably good approximate solutions for permutation-based optimization problems. 

\section{Ansatze for optimization problem-solving}\label{sec:p-circuit}

\subsection{Obtaining a doubly-stochastic matrix from a quantum circuit}

As mentioned in the introduction, some challenges of  \QAOA{} are related to the fact that constraints are not easily implementable and that it is complicated to embed any cost function as observable, especially in an amplitude encoding setting. We usually either have to prepare a specific state, or to derive a cost Hamiltonian and synthesize a circuit representing it. In this paper, we choose to do none of the technics mentioned above. Instead, we leverage a clever quantum circuit exploited recently by Mariella et al.~\cite{Mariella2024}.

\begin{figure}[ht]
\centering
\includegraphics[width=0.65\textwidth]{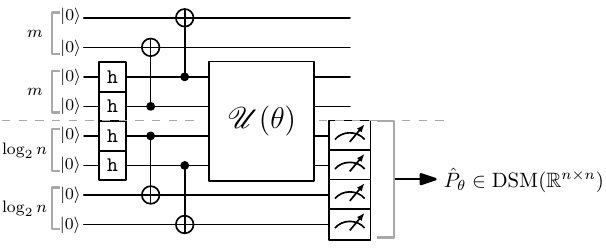}
\caption{Depiction of the variational circuit of Mariella et al.~\cite{Mariella2024} to obtain a doubly-stochastic matrix (DSM) starting from any unitary. For us this circuit will be denote by $\mathscr{P}(\theta)$. }
\label{fig:mariella}
\end{figure}

Consider the variational circuit presented in Figure~\ref{fig:mariella}, where $\mathscr{U}(\theta)\in \mathbb{C}^{2^m + n}$ is any arbitrary unitary matrix. The circuit is initialized in the $\ket{0}$ state. We have a total of $2 \log_2(n) + 2m$ qubits. The first represents the $n^2$ entries of a square matrix, while the additional $2m$ qubits are ancilla. We measure only the last $2 \log_2(n)$ qubits. 

Mariella and coauthors show in their Lemma~3.1 that, \emph{for any  $\mathscr{U}(\theta)$}, measuring the last $2 \log_2(n)$ qubits one obtains the corresponding entries of a $n\times n$ doubly-stochastic matrix. We recall that a doubly-stochastic matrix is a convex combination of permutation matrices. Furthermore, they also show that, if $m=0$, then the measurements deliver the entries of the matrix $\mathscr{U}(\theta) \odot \overline{\mathscr{U}(\theta)}$, where $\odot$ is the Hadamard entry-wise product and $\overline{(\cdot)}$ is the complex conjugate operator. In particular, if $m=0$ and $\mathscr{U}(\theta)$ is a permutation matrix, then the output of the circuit is the same permutation matrix. We summarize the results obtained in~\cite{Mariella2024} in the following proposition, for later use. 

\begin{proposition}{\cite{Mariella2024}}\label{prop.mariella} Let $\ket{i}$ refer to the $i$-th basis vector in the computational basis and let $
\ket{ij} = \ket{i} \otimes \ket{j}$. Define the \textit{row-major vectorization operator} on a matrix $M$ in $\mathbb{C}^{d\times d}$ as
\[
    \vecop(M):=\sum_{i=0}^{d - 1} M \ket{i} \otimes \ket{i}.
\]
Consider a bipartite Hilbert space $\mathcal{H} = \mathcal{H}_1 \otimes \mathcal{H}_2$, with dimensions $2^m$ and $n$, respectively. Recall that any (parametrized) unitary matrix $\mathscr{U}(\theta)\in \mathbb{C}^{2^m + n}$ can be decomposed into
\[
	\mathscr{U}(\theta) = \sum_{i=1}^{d^2} \lambda_i V_i(\theta) \otimes W_i(\theta),
\]
by the operator Schmidt decomposition, with $\{V_i\}$ and $\{W_i\}$ being sets of unitary operators orthogonal w.r.t. the Frobenius inner product, belonging to $\mathcal{H}_1$ and $\mathcal{H}_2$, respectively. Furthermore $\lambda_i \ge 0$, $\sum_i \lambda_i^2=1$, and $d^2 = \min\{(2^m)^2,n^2\}$. 

Consider now the circuit in Figure~\ref{fig:mariella}.  Partition the circuit into two subsystems. The first corresponding to the first auxiliary $2m$ qubits and the second corresponding to the last $2 \log_2(n)$ qubits. The mixed state corresponding the the last $2 \log_2(n)$ qubits is,
\begin{equation}
	\label{rho-def}
	\rho(\theta) = \frac{1}{n} \sum_{i=1}^{d^2} \lambda_i^2
	\vecop\left(W_i(\theta)\right) \vecop \left(W_i\right(\theta))^\dagger.
\end{equation} 
Furthermore, measuring the last $2 \log_2(n)$ qubits one obtains the following. Let, 
\begin{equation}\label{hadamard}
        p_{ij}(\theta) := n \mathrm{Tr}(\rho(\theta) \ket{ij}\bra{ij}) = \sum_{k=1}^{d^2} \lambda_k^2 \bra{i}\left(W_k(\theta) \odot \overline{W_k}(\theta)\right) \ket{j}
\end{equation}
for $i, j \in \intset{0}{n-1}$. Let $\left\{{\bf e}_{i}\right\}$ be the set of canonical basis vectors (with index $i$ starting from 0) for the vector space $\mathbb{R}^{n}$. Then the matrix
    \begin{align}
        \label{lemma:transp-plan-recovery-final-q}
        \hat{P}_{\theta}=&\sum_{i, j=0}^{n - 1} p_{ij}(\theta)\,{\bf e}_{i}{\bf e}_{j}^\top = \sum_{i, j=0}^{n - 1} p_{ij}(\theta)\ket{i}\bra{j},
    \end{align}
    is doubly stochastic.
\end{proposition}

Proposition~\ref{prop.mariella} tells us that given the density matrix $\rho(\theta)$ prepared as in Eq.~\eqref{rho-def}, the expectations w.r.t. the observables $\ket{ij}\bra{ij}$
provide the corresponding $(i, j)$ entry of $\hat{P}_{\theta}$, and this $\hat{P}_{\theta}$ is doubly stochastic (in fact, unistochastic). 

For $m=0$, and $\mathscr{U}(\theta) \in \Pi_n$, then $\hat{P}_{\theta} = \mathscr{U}(\theta) \odot \overline{\mathscr{U}(\theta)} = \mathscr{U}(\theta)$. Which is true, since the entries of a permutation matrix are only $0$ or $1$. 

For $m>0$, one can see the role of the auxiliary $m$ qubits, that is enlarging the function space as a result of the convex combination of density matrices in Eq.~\eqref{rho-def}. If $m=0$, then the number of terms in Eq.~\eqref{rho-def} reduces to 1, while for $m>0$, the number of is much larger (in fact, one can prove that all but two of the $\lambda_i$ must be positive~\cite{MuellerHermes2018}).

Having discussed how to generate doubly-stochastic matrices, we are now ready to see how to enlarge the number of explored permutations in our ansatze. 

\subsection{Increasing the number of explored permutations}

As we have observed in Theorem~\ref{th:number-of-permutations}, the number of spanned permutation scales as $2^{O(q^2)}$ if $q=2^n$ is the number of qubits used to represent a permutation matrix of dimension $n$. This is the best we can do with local gates, but not quite close to the total number of permutations $n!$, as we can see on Table~\ref{table: span}. 

The idea is now to enlarge the spanned space by adding extra qubits as in the circuit in Figure~\ref{fig:mariella}, with the unitaries that we have constructed in Section~\ref{sec:circuits}. With this, we can obtain the following advantages. 

First, by Eq.~\eqref{rho-def}, we can see that the number of convex combination augments, giving us the hope that adding addition qubits is similar to spanning a much larger space and then projecting it onto our problem space $n$. This hope is supported by theory (below) and simulation studies. 

Second, since to compute the optimization cost $f(P)$ will typically involve $O(n^2)$ arithmetic operations as well as storage, a unitary in $\mathbb{C}^{n\times n}$ is classically simulatable by the same computer. By increasing the number of qubits and choosing $m \gg n$, then our unitaries will cease to be classically simulatable (even though the cost will remain so) and the algorithm derived upon them will be genuinely quantum. 

To support the first advantage, we can now prove a useful theoretical result, which characterizes the output $\hat{P}_{\theta}$, whenever $\mathscr{U}(\theta)$ is a permutation matrix. Specifically, we choose the Bruhat unitary, $\mathscr{U}^{\mathrm{Bruhat}}(\theta)$, and the Borel unitary, $\mathscr{U}^{\mathrm{Borel}}(\theta)$, both obtained in Section~\ref{sec:circuits}, and reported in Figure~\ref{fig:ansatze}, for simplicity.  

\begin{proposition}\label{prop.decomps}
Consider the circuit in Figure~\ref{fig:mariella}, with unitaries $\mathscr{U}(\theta)$ taken as $\mathscr{U}^{\mathrm{Borel}}(\theta)$, or $\mathscr{U}^{\mathrm{Bruhat}}(\theta)$ . Let $\theta \in \{0,\pi\}^\ell$ for all the $\ell$ parameters, so that the unitary represent a permutation matrix and recall $n=2^q$. Then, 
$$
\hat{P}_{\theta} = \sum_{i=1}^w \lambda_i P_i, \quad P_i \in \mathrm{span}(\mathrm{Bruhat}_n),
$$
with $w>1$ whenever $m>0$, i.e., $\hat{P}_{\theta}$ is a convex combination of permutation matrices belonging to the set of permutations in the Bruhat span of dimension $n$.  In addition,
\begin{eqnarray*}
\mathscr{U}^{\mathrm{Bruhat}}(\theta)&:& \, w\leq \max\{2^{3 m q} , |\mathrm{span}(\mathrm{Bruhat}_n)|\}; \\ \mathscr{U}^{\mathrm{Borel}}(\theta)&:& \, w\leq \max\{2^{m q} , |\mathrm{span}(\mathrm{Bruhat}_n)|\}.
\end{eqnarray*}
Finally, the number of distinct $\hat{P}_{\theta}$ is at most equal to the cardinality of the Bruhat span of dimension $n 2^m = 2^{q+m}$. 
\end{proposition}

\begin{proof}
The proof is based on the operator Schmidt decomposition of \texttt{cx} and \texttt{swap} gates, which are the gates that connects the $m$ part to the $n$ part of the circuit. Whenever there is a connecting gate, either the parameter is $0$, and the gate is the identity, or the parameter is $\pi$ and we have a \texttt{cx} or a \texttt{swap} depending on the parametrized connecting gate. Up to a renormalization, we have 
\[
\mathtt{cx} \simeq \ket{0}\bra{0} \otimes I + \ket{1}\bra{1} \otimes X, \qquad \text{and} \qquad \mathtt{swap} \simeq I\otimes I + X\otimes X + Y\otimes Y + Z\otimes Z,
\]
see for instance~\cite{Coffey2008}. Since the decomposition involves only matrices which have one non-zero elements per row and column, whose modulus is $1$, then, $W_i(\theta) \odot \overline{W_i}(\theta)$ in Eq.~\eqref{hadamard} is a permutation matrix obtained combining local gates. Specifically, the $Y$'s will act as $X$ and the $Z$'s act as $I$ in the SWAP gate. This implies that the number of distinct $W_i(\theta) \odot \overline{W_i}(\theta)$ matrices is upper bounded by the span of the cardinality of the Bruhat span of dimension $n$. The number of terms in the convex combination is also upper bounded by the number of splits caused by connecting gates. The number of connecting gates are $3mq$ for Bruhat and $mq$ for Borel, giving an upper bound of $2^{3mq}$ and $2^{mq}$, respectively. 

Finally, any locally constructed permutation $\mathscr{U}(\theta)$ in $2^{q+m}$ dimension will be among the ones in the Bruhat permutation set of dimension $n\ 2^m$. Each of them will give rise to a different operator Schmidt decomposition and henceforth we can count the maximum number of distinct $\hat{P}_{\theta}$. 
\end{proof}

A few remarks are in order. 

\begin{remark} Whenever an ancilla is used, $\hat{P}_{\theta}$ will be a doubly-stochastic matrix, even if $\mathscr{U}(\theta)$ was a permutation matrix in $2^m + n$ (i.e., all the $\theta$'s were either $0$ or $\pi$). This is not a problem, since we know how to project onto the permutation matrices classically (see below and~\cite{Fogel2015, Barvinok2006}). 
\end{remark}

\begin{remark}
Due to the number of unique $\hat{P}_{\theta}$, whenever $\mathscr{U}$ is a permutation, one can compute a lower bound on the number of required ancilla qubits (imagining that each doubly-stochastic matrix maps into a unique permutation) to span the set of permutations acting $n$ symbols, which gives us 
$$
2^{(m + \log_2(n))^2} \approx n! \implies m \gtrsim \Omega(\sqrt{n \log n}).  
$$
This lower bound is quadratically lower than the number of qubits required in \QAOA{}~\cite{Glos2022}. In addition, here we have the choice of selecting how much of the set we span by augmenting or reducing the number of used ancillas. This tuning knob gives us the power to trade-off span and implementability on current hardware, which has no parallel in  \QAOA{}. 
\end{remark}

\begin{remark}
Due to the presence of the operator Schmidt decomposition characterizing the exact span of $\hat{P}_\theta$ is challenging to do, but the interested reader can see some interesting works in this respect~\cite{Chen2016,Benoist2017,MuellerHermes2018}. 
\end{remark}

\subsection{Considered $\mathscr{U}(\theta)$ and empirical quantum boost}

\begin{figure}[ht]
\centering
\includegraphics[width=0.9\textwidth]{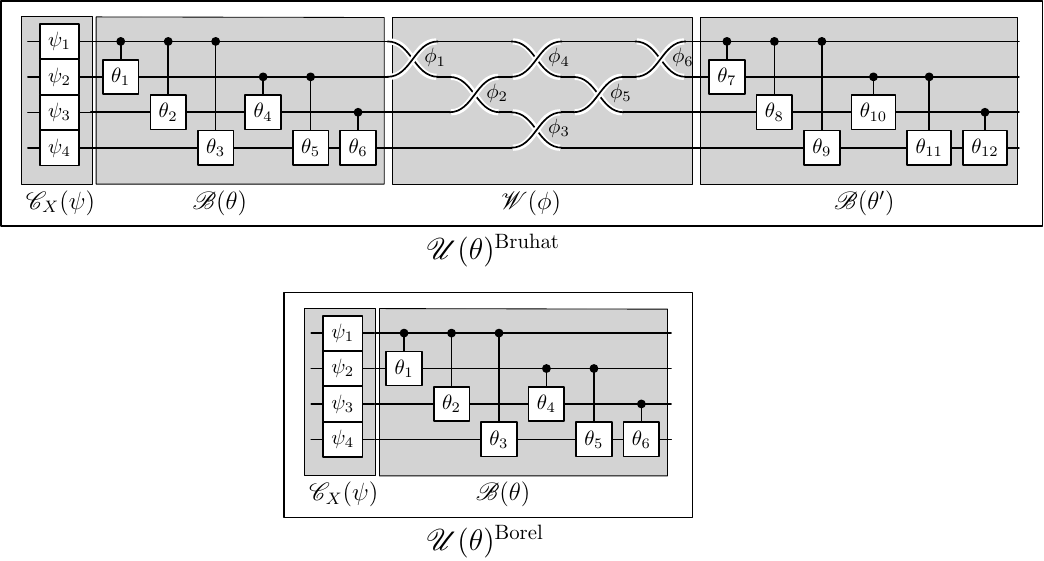}
\caption{Depiction of the proposed ansatze for four qubits. With a slight abuse of notation, in both $\mathscr{U}(\theta)$, the parameters $\theta$ collect all the parameters in the ansatz. For example for Bruhat, $\theta := \{\psi, \theta, \phi, \theta'\}$. Single qubit gates with one angle indicate $\mathtt{rx}$ gates.  }
\label{fig:ansatze}
\end{figure}

We can now explore by numerical simulations the total number of unique permutations generated by our quantum circuit (Figure~\ref{fig:mariella}) with a few different unitaries $\mathscr{U}(\theta)$. For comparison, we use the Bruhat unitary, $\mathscr{U}^{\mathrm{Bruhat}}(\theta)$, and the Borel unitary, $\mathscr{U}^{\mathrm{Borel}}(\theta)$, both obtained in Section~\ref{sec:circuits}, and reported here in Figure~\ref{fig:ansatze}. 

The Borel unitary is explored here since, for some applications, it could be beneficial to have a weaker circuit, spanning less permutations, but in return that requires less classical parameters, and have a shallower depth relatively to the number of parameters.

We also use a widely used unitary that has been proposed in the context of quantum machine learning: the Strong Entangling Layer (SEL) unitary~\cite{Schuld2020} as baseline. 

\begin{figure}[ht]
\centering
\includegraphics[width=0.95\textwidth]{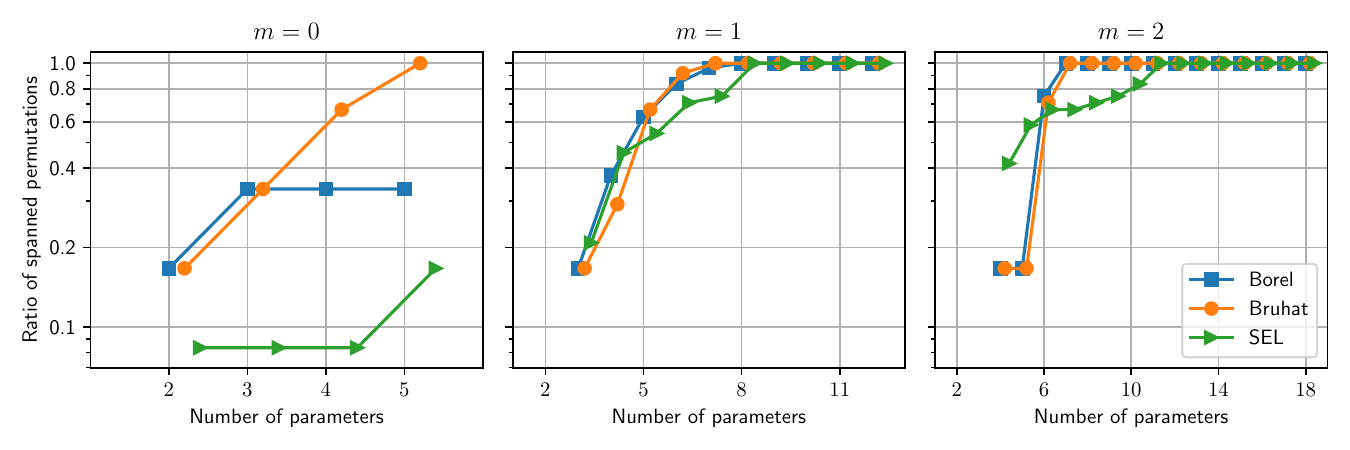}
\caption{Ratio of the spanned permutations as a function of the number of parameters for $q=2$, i.e., $n=4$. The lines are shifted slightly horizontally for better readability. Recall that the Bruhat span of $n$ is $24$ and $n!=24$.}
\label{fig.span2}
\end{figure}

\begin{figure}[ht]
\centering
\includegraphics[width=0.95\textwidth]{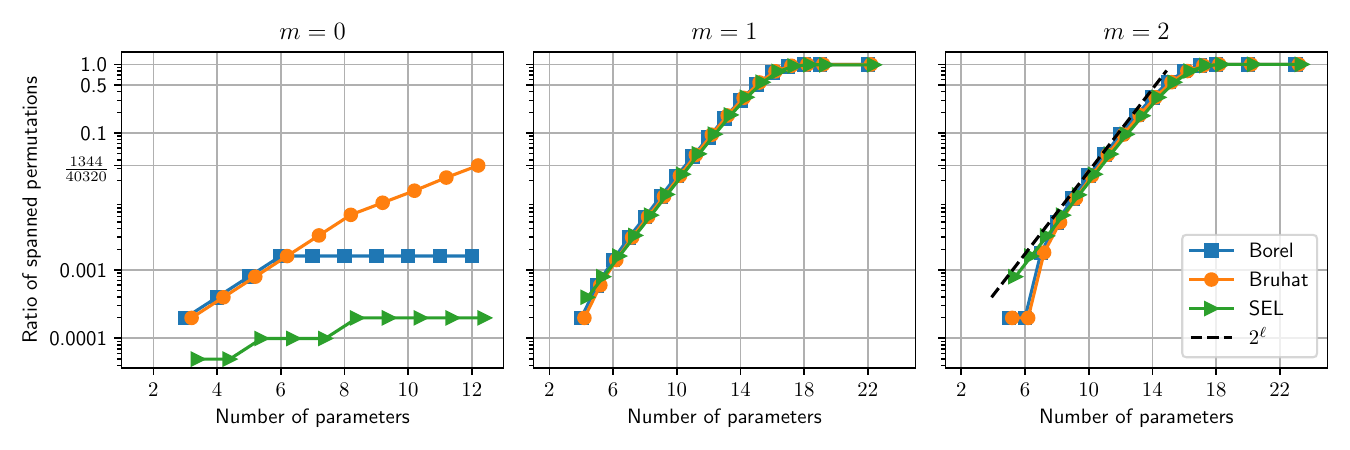}
\caption{Ratio of the spanned permutations as a function of the number of parameters for $q=3$, i.e., $n=8$. The lines are shifted slightly horizontally for better readability. Recall that the Bruhat span of $n$ is $1,344$, $n!=40,320$, while the Bruhat span of $n \times 2^1$ is $322,560$.}
\label{fig.span3}
\end{figure}

We report in Figures~\ref{fig.span2} and \ref{fig.span3} the obtained spans performing one projection into the permutation set if $m>0$. Both figures feature the number of spanned permutations depending on the number of parameters that we have in the circuit. For each number of parameters $\ell$, we consider either all the possible combinations of the parameters in $\{0, \pi\}$, i.e., $2^\ell$, or $250,000$ samples among them, whichever is the lowest. As such, we consider discrete parameters: this fits the presented theory and the practical results, as we introduce regularization terms in the optimization problem to force the parameters in $\{0, \pi\}$. 

As we can observe in the figures, for $m=0$, the Bruhat ansatz performs the best, obtaining the best we can achieve. For $m>0$, all the ansatze perform similarly and offer a significant boost in spanned permutations, arriving close\footnote{The larger span for $m=1$ is obtained by Bruhat at $40,199$ and for $m=2$ by Borel and SEL at $40,243$ with a Bruhat at $40,234$.} to the total number of permutation $8! = 40,320$. Interestingly, the boost in performance, which we could call as quantum boost, since for $m>0$ we are in a genuine quantum regime, is nearly identical for $m=1$ and $m=2$ and depends quite closely on the number of parameters as $2^\ell$ (i.e., the number of distinct possibilities). As such, for the same number of parameters $\ell$, quantum circuits with greater $m$'s will be shorter than the ones with lower $m$'s. 

\begin{figure}[t]
\centering
\includegraphics[width=0.45\textwidth]{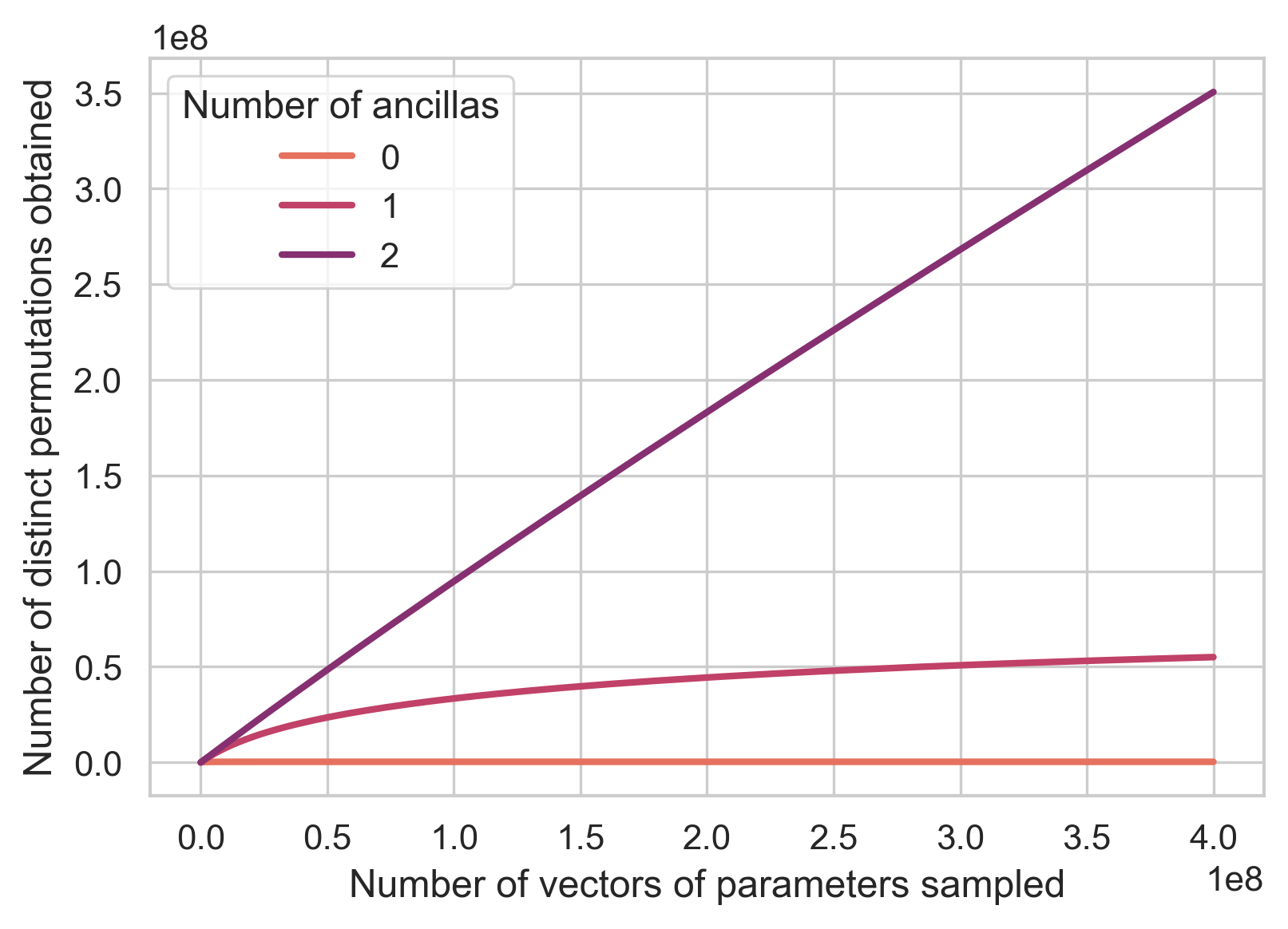}
\includegraphics[width=0.45\textwidth]{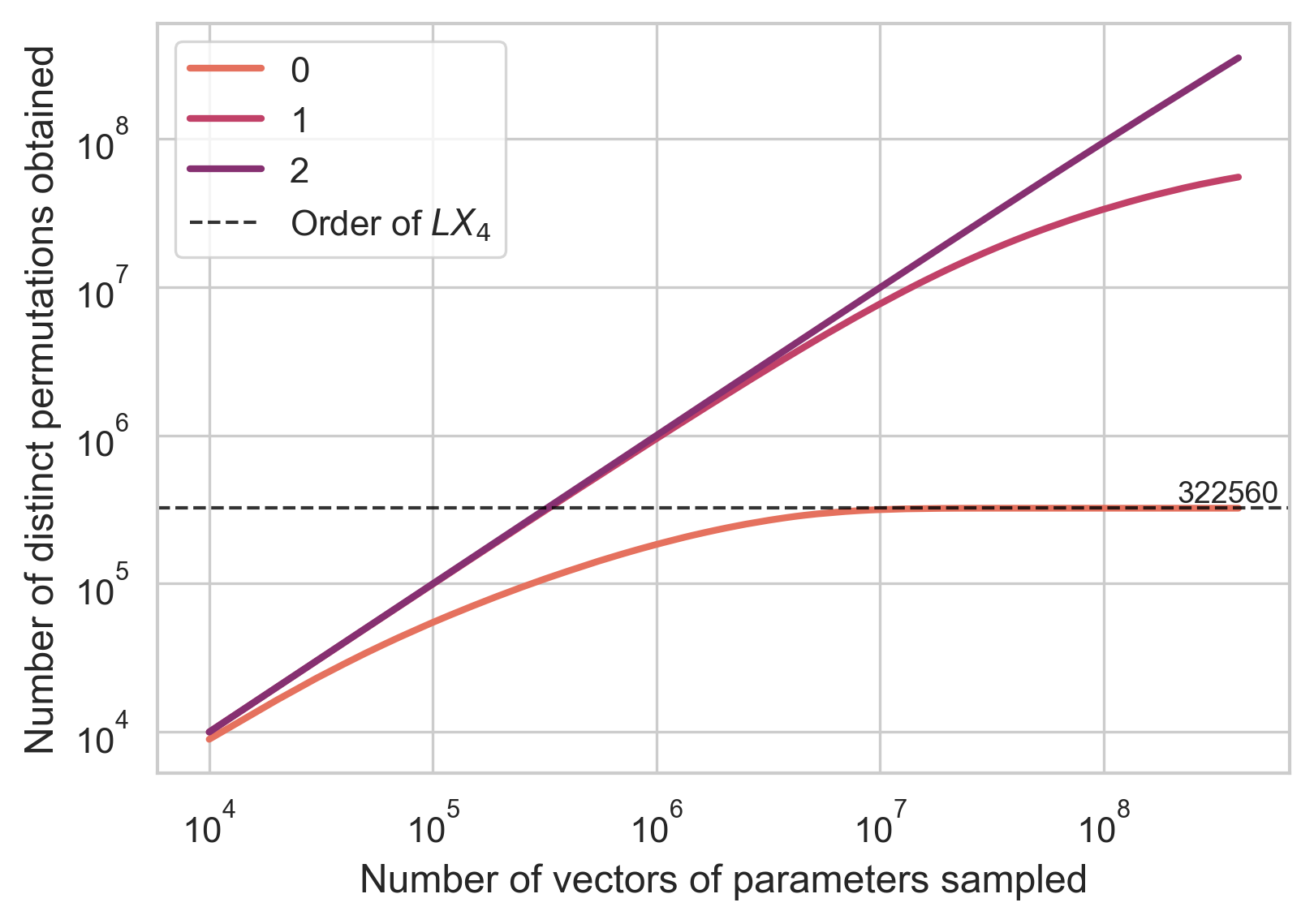}
\caption{Number of permutations spanned as a function of the number of vectors of parameters sampled for \(q=4\), i.e., \(n=16\), plotted using a standard and a log-log scaling. The dashed horizontal line indicates the order of the group \(LX_4\), i.e., the Bruhat span for \(q=4\) and \(n=16\).}
\label{fig.span4}
\end{figure}

The similarities of the behavior for different $m$'s is due to the low number of considered qubits for $n=8$. As discussed in Proposition~\ref{prop.decomps}, we know that the number of spanned permutation will be at most,
$$
\min \Big\{ 2^\ell, 2^{\frac{q'(q'+1)}{2}} \prod_{k=1}^{q'} (2^k-1), n! \Big\}, \quad q' = \log_2(n)+m,
$$
thereby indicating a difference in performance for different $m$'s starting from $n=16$. To test this, we run the Bruhat's ansatz with \(q=4\) and $m=0, 1, 2$ over $400$ millions different parameterizations\footnote{Since $\theta \in \{0, \pi\}^{\ell}$, there are $2^\ell$ distinct parameterizations (\(\sim10^6\) for \(m=0\), \(\sim10^{10}\) for \(m=1\) and \(\sim10^{15}\) for \(m=2\)) and $\sim 2 \times 10^{13}$ possible permutations. There are too many of them to explore exhaustively as we did for \(2\) and \(3\) qubits.} of $\theta$, and counted the number of distinct permutations returned. We observe in Figure~\ref{fig.span4} that increasing \(m\) significantly improves the span of the circuit by one or two orders of magnitude, thereby empirically validating our approach. 

\section{Application to optimization problems}\label{sec:optimiz}

We are now ready to see how to use our quantum circuit to the resolution of permutation-based optimization problems as Problem~\eqref{eq:perm:prob}. We will focus on quadratic assignment problems and graph isomorphism.

In all the problems, we will use the \ADAM{} optimizer to optimize classically for the $\theta$. We will also use a projection-onto-the-permutations technique discussed in Appendix~\ref{ap.permutation}.

\subsection{Quadratic assignment problem}

The first optimization problem we consider is the quadratic assignment problem~\cite{koopmans1957assignment} (or QAP for short), for which we can set in Problem~\eqref{eq:perm:prob}, the objective function \(f\) given by
$$
f(P): = \mathrm{tr} \left( W P D^\top P^\top \right),
$$
where $W$ and $D$ are two matrices in $\mathbb{R}^{n \times n}$. While there exist more general formulations of the problem (as well as polynomial generalization)~\cite{lawler1963quadratic,Pardalos1999}, the cost we have selected is general enough to consider most of the interesting instances appeared in the literature. 

The quadratic assignment problem is an important problem in combinatorial optimization, with widespread use in practice and whose research stays very active~\cite{Tan2024}. For example, the travel salesperson problem is an instance of the QAP. Moreover, the QAP is \textsf{NP}-hard, and even finding an \(\varepsilon\)-approximation is intractable. 

\begin{theorem}[Sahni and Gonzalez~\cite{sahni1976p, burkard1998quadratic}]
The quadratic assignment problem is strongly \(\mathsf{NP}\)-hard. Moreover, for an any \(\varepsilon > 0\), the existence of a polynomial time \(\varepsilon\)-approximation algorithm for the quadratic assignment problem implies \(\mathsf{P} = \mathsf{NP}\). 
\end{theorem}

It was later shown by Pardalos, Rendl and Wolkowicz~\cite{pardalos1994quadratic} that even finding a locally optimal solution, relatively to some specific neighborhood structure, is among the hardest local search problems. For a comprehensive examination of the quadratic assignment problem, see the survey of Burkard, Çela, Pardalos and Pitsoulis~\cite{burkard1998quadratic}. 

\subsection{Quantum QAP solver: \Algo{}}

We implement our code in python and use the package Pennylane~\cite{bergholm2018pennylane} for quantum computations. All the circuits are simulated exactly. We have implemented three different ansatze, the Borel, the Bruhat, and the SEL. 

As depicted in Figure~\ref{fig.approach}, we can think of the whole quantum circuit as a mapping \(\theta \mapsto \hat{P}_\theta\), with \(\hat{P}_\theta\) being a doubly stochastic matrix. We therefore aim at minimizing the following objective function
\[
\ell(\theta) \coloneqq f(\hat{P}_\theta) = \mathrm{tr} \left( W \hat{P}_\theta D^\top \hat{P}^\top_\theta \right), 
\]
which is a non-standard non-convex continuous relaxation of the objective function of quadratic assignment problems. To guide the circuit to return doubly stochastic matrices that are closed to permutations, we regularize this objective function, which becomes
\[
\ell_R(\theta) = \ell ( \theta ) + \frac{1}{10} \mathrm{st}(\hat{P}_\theta) + \frac{15}{100} S_\varepsilon (\hat{P}_\theta) + \frac{1}{100} \mathrm{ort}(\hat{P}_\theta)
\]
with
\[ \begin{cases} 
\mathrm{st}(\hat{P}_\theta) & = \sum_{j = 1}^n \left( \sum_{i = 1}^n \left[ \hat{P}_\theta \right]_{ij} - 1 \right)^2, \\
S_\varepsilon (\hat{P}_\theta) & = - \sum_{i, j = 1}^n \left[ \hat{P}_\theta \log \left( \hat{P}_\theta + \varepsilon \right) \right]_{ij}, \\
\mathrm{ort} (\hat{P}_\theta) & = \sum_{i, j = 1}^n \left(\hat{P}_\theta^\top \hat{P}_\theta - I \right)^2.
\end{cases} \]
Each additional term helps at forcing \(\hat{P}_\theta\) to be a permutation, and at smoothing the loss landscape of the objective function.


We search for a minimum of \(\ell_R\) using \textsc{Adam} with Nesterov momentum, so that the vector of parameters \(\theta^{(k)}\) at step \(k\) is updated according to the following rule 
\[
\theta^{(k + 1)} = \theta^{(k)} - \eta_k \frac{\overline{\mu}^{(k)}}{\sqrt{\overline{\nu}^{(k)}} + \varepsilon} \quad \text{with} \quad \begin{cases}
\mu^{(k)} = \beta_1 \mu^{(k-1)} + (1 - \beta_1) g_k, \\
\nu^{(k)} = \beta_2 \nu^{(k-1)} + (1 - \beta_2) g_k^2, 
\end{cases}
\]
\[
\text{as well as} \quad \overline{\mu}^{(k)} = \frac{\beta_1}{1 - \beta_1^{k+1}} \mu^{(k)} + \frac{1 - \beta_1}{1 - \beta_1^k} g_t \quad \text{and} \quad \overline{\nu}^{(k)} = \frac{\nu^{(k)}}{1 - \beta_2^k},
\]
with \(g_k = \nabla_\theta \ell_R ( \theta^{(k-1)} )\) and the usual hyperparameters fixed at 
\[
\eta = 0.005, \quad \beta_1 = 0.9, \quad \beta_2 = 0.999, \quad \text{and} \quad \varepsilon = 10^{-8}.
\]
At every iteration we transform the double stochastic matrix \(\hat{P}_\theta\) in two different ways:
\begin{itemize}
\item by solving a linear assignment problem using the Hungarian algorithm,
\item by taking the permutation that alters the order of the entries of a randomly drawn vector the same way as the double stochastic matrix \(\hat{P}_\theta\), as explained in Appendix~\ref{ap.permutation}.
\end{itemize}
We then evaluate the original loss function \(\ell\) at both permutations, keep the minimum of the two, and then compare with the best permutation currently known.

We initialize the parameters $\theta$ by drawing them around the center of the span with some uniform distribution. Since the ansatz with a given ancilla $m>0$ contains the ansatz with ancilla $m-1$, we initialize the parameters $\theta$ of $m$ with the ones of $m-1$ whenever they are accessible (this last step is not necessary, but it helps having consistent results). 

The resulting algorithm is presented in Algorithm~\ref{algo:qap} and is called \Algo{}.

\begin{algorithm}[H]
\caption{\Algo{}}\label{algo:qap}
\begin{algorithmic}[1]
\Require Ansatz, number of ancilla qubit $m^*$, number of iterations $I$
\State Define \(\theta^{(0)}\) uniformly in \(\frac{\pi}{2} \pm 0.05\) according to the chosen ansatz
\For{\(m\) in \(\{0, \ldots, m^*\}\)}
\While{convergence not reached}\Comment{We set \(I=1000\) iterations}
\State \( \theta^{(k+1)} \gets \Call{Adam}{\theta^{(k)}} \) \Comment{\ADAM{} calls the quantum circuit}
\State get $\tilde{P}$ by projecting $\hat{P}_{ \theta^{(k+1)}}$ following Appendix~\ref{ap.permutation}
\State get \(\tilde{v}\) by evaluating the cost classically
\If{\(\tilde{v} < v\)}
\State \(v \gets \tilde{v}\), \(P \gets \tilde{P}\)
\EndIf
\EndWhile
\State \textbf{return} \(v\), \(P\)\Comment{We returned the results for each \(m\)}
\State pad the final \(\theta\) with \(\frac{\pi}{8}\) for the gates that will be added with the increment of \(m\)
\EndFor
\end{algorithmic}
\end{algorithm}

\subsubsection{Numerical results: QAPlib}

We start by testing \Algo{} on instances of QAP collected in the QAPlib benchmark repository~\cite{Burkard1997}. The instances range from small scale $n=16$ to large scale $n=256$ and beyond, the largest having $n=729$ vertices. The instances are known to be hard, and for some of them we do not know the optimal permutation value. We consider all the solved instances in the benchmark for the following number of nodes $n \in \{16, 32, 64, 128, 256\}$, which have $\{12, 7, 2, 1, 1\}$ instances, respectively. 

We compare \Algo{} with two baselines. First, we use a state-of-the-art classical heuristic~\cite{Vogelstein2015}, which is implemented in Python via SciPy. The heuristic solves a sequential convexification of the quadratic assignment problem, by relaxing the permutation constraint into a double stochastic matrix constraint and then using the Frank-Wolfe algorithm to solve the resulting continuous convex problem. Then they would project onto the permutation set by a linear assignment problem. In this context, the heuristic of~\cite{Vogelstein2015} is close in spirit to our quantum heuristic, even though we span the set differently\footnote{We notice that the classical heuristic in~\cite{Vogelstein2015} is not necessarily the best on every problem, other exists in the literature, but it is quite robust, it can scale quite well and serves to us as a baseline. }. This algorithm has an empirical complexity of \(O(n^3)\), with small leading constants (\(\sim 10^{-9}\)).

Second, we implement a naive random method, which draws permutations randomly. The number of random trials is set to be equal to the number of random projections we perform in \Algo{}: $50 \lceil \frac{I}{10} \rceil$, which is generally small compared to the permutation set.

\begin{figure}[ht]
\centering
\includegraphics[width=0.95\textwidth]{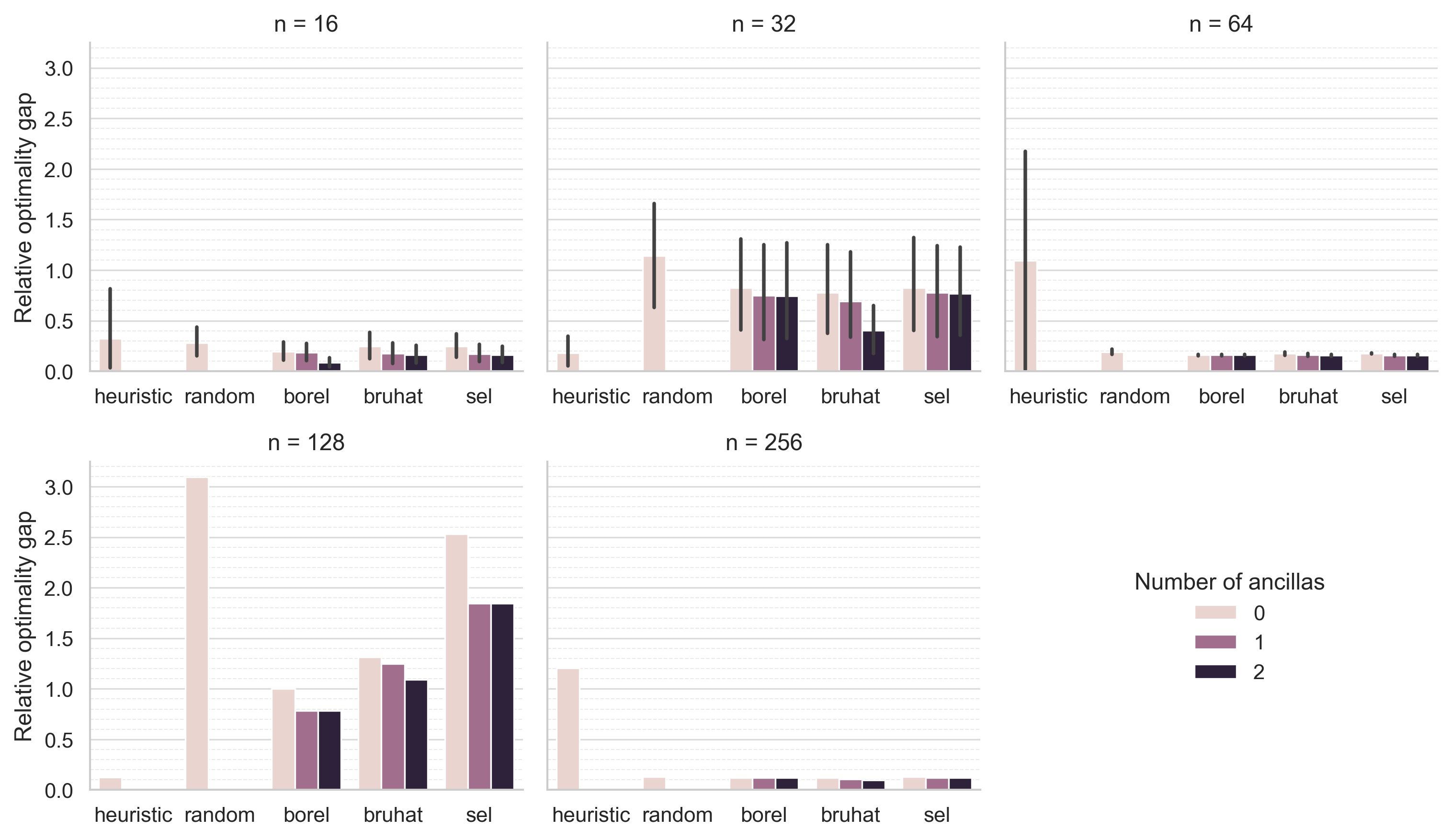}
\caption{Numerical comparisons on instances taken from QAPlib. `heuristic' denotes the classical heuristic~\cite{Vogelstein2015} and `random' the naive random method. We present here the mean over the instances and the \(95\%\) confidence interval. The full results are available in Appendix~\ref{ap.qapres}. }
\label{fig.QAPlib}
\end{figure}

\begin{figure}[H]
\centering
\includegraphics[width=0.95\textwidth]{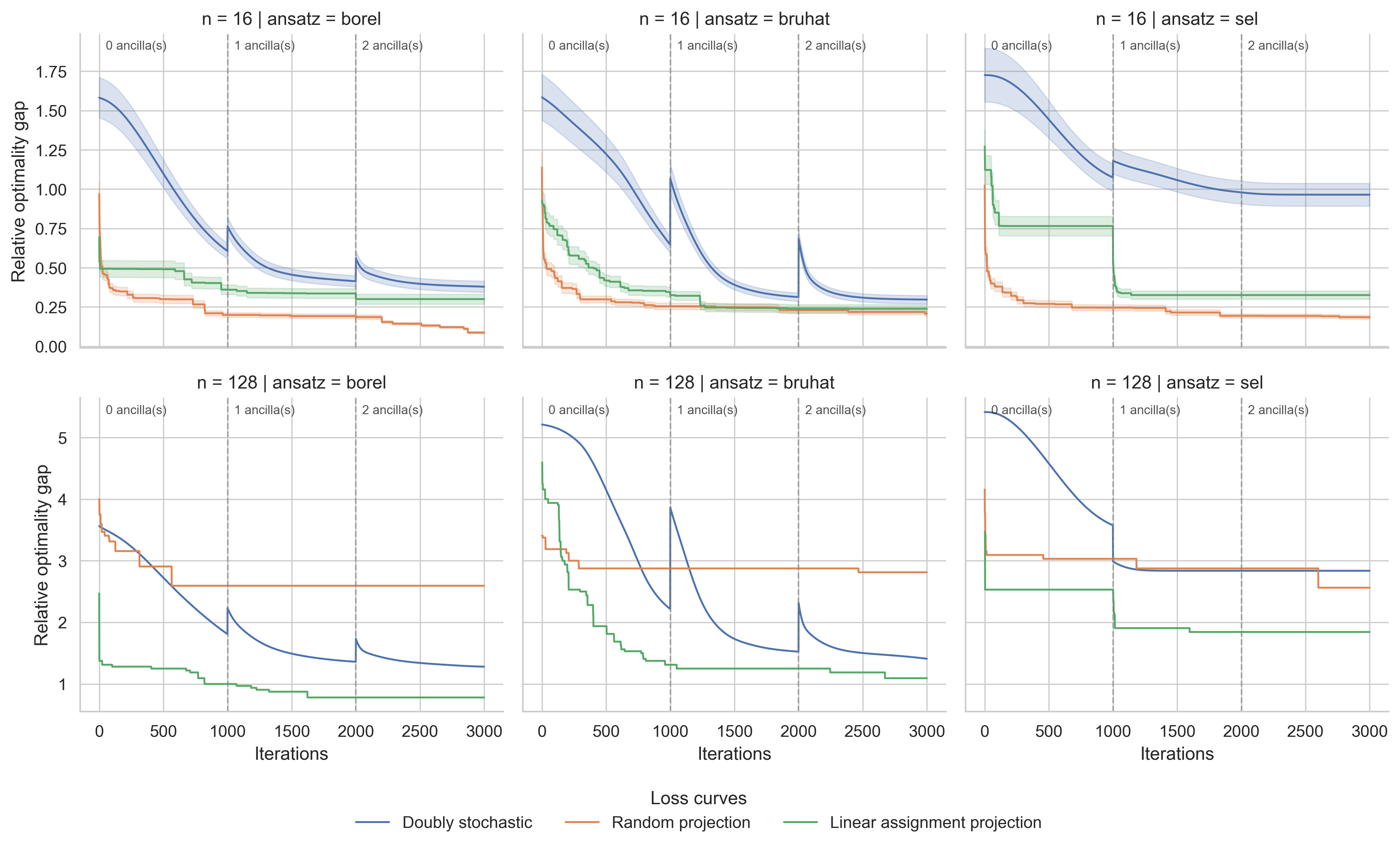}
\caption{Numerical comparison on instances taken from QAPlib of the value of the loss functions, and the cost associated to the two different permutations obtained from the current doubly stochastic matrix returned by the circuit. Vertical dashed lines indicate the iterations at which an additional ancilla qubit is introduced in the variational ansatz. The full results are available in Appendix~\ref{ap.qapres}.}
\label{fig.convergence}
\end{figure}

In Figure~\ref{fig.QAPlib} and Figure~\ref{fig.convergence}, we report the results for our algorithm in different settings as well as the other baselines. The results are reported in terms of the relative optimality gap, defined as,
\[
\textrm{relative optimality gap} = \frac{f(\tilde{P})-f^*}{f^*},
\]
since for these instances we know the optimal solution value $f^*$. 

As we can see, \Algo{} achieves good performance compared to the classical heuristic~\cite{Vogelstein2015} and it is better than the naive random method. For some instances, \Algo{} outperforms the classical algorithm ($n= \{16, 64, 256\}$), for the others, the classical algorithm appears to be better. We can also see the quantum benefit of using a $m>0$ (since the ansatz at $m+1$ includes the one at $m$, a greater $m$ is always not worse, provided the classical optimizer can optimize the parameters exactly). The choice of ansatz seems to have a smaller influence on the result, but in general we think this is mainly due to the difficulty of the classical optimizer to optimize, rather than the expressivity of the ansatz. 

Figure~\ref{fig.QAPlib} and the accompanying Appendix~\ref{ap.qapres} depict a quantum algorithm (\Algo{}) which can tackle large instances, it has good potential, and it could be in par with classical heuristics on difficult QAP instances. For reference, the largest circuit that we have used for the case $n=256$, $m=2$ has $20$ qubits. Interesting future work here would be to bring \Algo{} onto HPC architectures and simulate large instances with large $m$, where we expect a clearer quantum benefit. Also, for large $m$, we may experience a more favorable optimization landscape, due to over-parametrization.

\subsubsection{Numerical results: Random instances}

We move on to evaluate \Algo{} on random instances, generated by drawing the elements of matrices $W$ and $D$ uniformly between \(0\) and \(10\). 

\begin{figure}[H]
\centering
\includegraphics[width=0.99\textwidth]{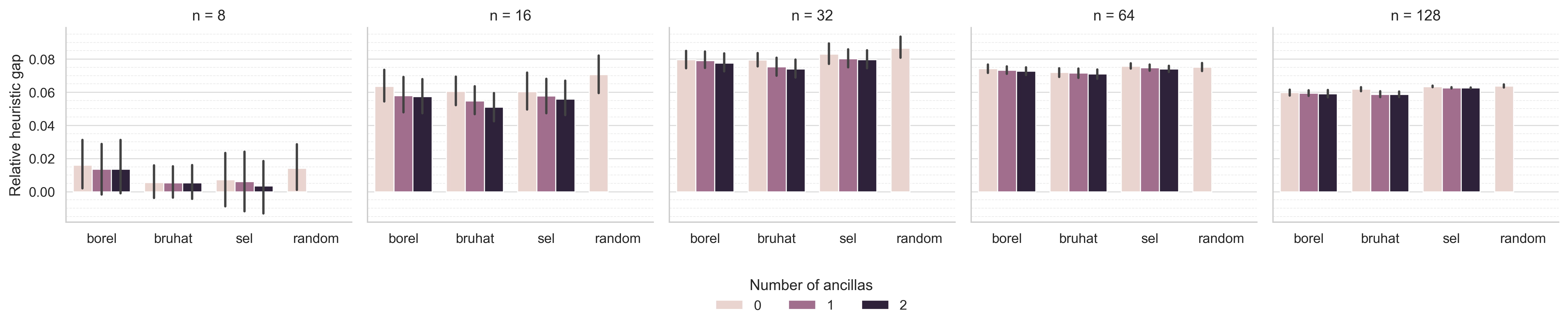}
\caption{Numerical comparisons on $10$ uniformly drawn random instances (\(5\) for n = 128) with respect to the classical heuristic~\cite{Vogelstein2015}. We present here the mean and the 95\% confidence interval. }
\label{fig.QAP.rand}
\end{figure}

We compare the naive random method and \Algo{} with the Borel ansatz as a percentage difference with respect to the classical heuristic. We report the results in Figure~\ref{fig.QAP.rand}. As we can see,  \Algo{} is competitive on random instances and it is generally better than the classical heuristic for $n=8$. We believe that the fact that the performance of \Algo{} degrades for larger instances is mainly due to the classical optimizer, which fails to find a better minimum.  

\subsection{Graph isomorphism problem}

The second optimization problem we study is the graph isomorphism problem~\cite{Bollobas1998},  for which we can set in Problem~\eqref{eq:perm:prob}, the objective function \(f\) given by
$$
f(P): = \lVert A - P B P^\top \rVert^2_{\mathrm{F}},
$$
where $A, B \in \{0,1\}^{n\times n}$ are the adjacency matrix of two undirected graphs $\mathcal{A}$ and $\mathcal{B}$, respectively, both with $n$ vertices and $\lVert \cdot \rVert_{\mathrm{F}}$ is the Frobenius norm. An example of a graph isomorphism problem instance is given in Figure~\ref{fig.instance}. We remark here that we could already tackle the harder sub-graph isomorphism problem with a similar formulation, but we focus on the easier graph isomorphism for simplicity. 

\begin{figure}[H]
\begin{center}
\begin{tikzpicture}[shorten >=1pt,->]
  \tikzstyle{vertex}=[circle, fill=black!25, minimum size=12pt, inner sep=2pt]
  \node[vertex] (G_1) at (-1,- 0.5) {1};
  \node[vertex] (G_2) at (0, 0)  {2};
  \node[vertex] (G_3) at (1, -0.5)  {3};
  \node[vertex] (G_4) at (1.5, -1.5) {4};
  \node[vertex] (G_5) at (1, -2.5) {5};
  \node[vertex] (G_6) at (0, -3) {6};
  \node[vertex] (G_7) at (-1, -2.5) {7};
  \node[vertex] (G_0) at (-1.5, -1.5) {0};  
  \draw (G_0) -- (G_4) -- cycle;
  \draw (G_0) -- (G_5) -- cycle;
  \draw (G_0) -- (G_7) -- cycle;
  \draw (G_1) -- (G_2) -- cycle;
  \draw (G_1) -- (G_3) -- cycle;
  \draw (G_1) -- (G_4) -- cycle;
  \draw (G_1) -- (G_5) -- cycle;
  \draw (G_1) -- (G_6) -- cycle;
  \draw (G_1) -- (G_7) -- cycle;
  \draw (G_2) -- (G_5) -- cycle;
  \draw (G_2) -- (G_6) -- cycle;
  \draw (G_2) -- (G_7) -- cycle;
  \draw (G_3) -- (G_6) -- cycle;
  \draw (G_3) -- (G_7) -- cycle;
  \draw (G_4) -- (G_5) -- cycle;
  \draw (G_4) -- (G_6) -- cycle;
  \draw (G_5) -- (G_6) -- cycle;
\end{tikzpicture}
\hspace{2cm}
\begin{tikzpicture}[shorten >=1pt,->]
  \tikzstyle{vertex}=[circle,fill=black!25,minimum size=12pt,inner sep=2pt]
  \node[vertex] (G_1) at (-1,- 0.5) {1};
  \node[vertex] (G_2) at (0, 0)  {2};
  \node[vertex] (G_3) at (1, -0.5)  {3};
  \node[vertex] (G_4) at (1.5, -1.5) {4};
  \node[vertex] (G_5) at (1, -2.5) {5};
  \node[vertex] (G_6) at (0, -3) {6};
  \node[vertex] (G_7) at (-1, -2.5) {7};
  \node[vertex] (G_0) at (-1.5, -1.5) {0};  
  \draw (G_0) -- (G_2) -- cycle;
  \draw (G_0) -- (G_5) -- cycle;
  \draw (G_0) -- (G_6) -- cycle;
  \draw (G_1) -- (G_3) -- cycle;
  \draw (G_1) -- (G_4) -- cycle;
  \draw (G_1) -- (G_5) -- cycle;
  \draw (G_1) -- (G_6) -- cycle;
  \draw (G_2) -- (G_4) -- cycle;
  \draw (G_2) -- (G_6) -- cycle;
  \draw (G_2) -- (G_7) -- cycle;
  \draw (G_3) -- (G_4) -- cycle;
  \draw (G_3) -- (G_5) -- cycle;
  \draw (G_3) -- (G_6) -- cycle;
  \draw (G_3) -- (G_7) -- cycle;
  \draw (G_5) -- (G_6) -- cycle;
  \draw (G_5) -- (G_7) -- cycle;
  \draw (G_6) -- (G_7) -- cycle;
\end{tikzpicture}
\end{center}
\caption{Two isomorphic graphs with $P = (4 \; 6 \; 7 \; 0 \; 1 \; 3 \; 5 \; 2)$.}\label{fig.instance}
\end{figure}

The graph isomorphism problem has numerous applications when data can be represented as networks~\cite{CONTE2004,Bonnici2013,Lee2012}. Mathematically, the graph isomorphism problem is not known to be \textsf{NP}-hard in general~\cite{Arvind2000}, even though is still hard to solve in practice. However, many classical algorithms exist tailored to different real-word graphs and instances that can deal
with graphs with thousands to millions of vertices and edges~\cite{Cordella2004,Lee2012,Aflalo2015,Nabti2018}.

We choose to focus here on the graph isomorphism to provide a quantum comparison of our approach to the amplitude encoding methodology presented in~\cite{Mariella2023}, which we expect to be somewhat of similar performance. It is also worth noticing that graph problems are reasonable to encode on photonic quantum computers~\cite{Mezher2023}, thereby making a good benchmark for future comparison of different technologies. Finally, we remark that a  \QAOA{} approach to this problem, tempted in~\cite{Calude2017}, requires $O(n^2)$ qubits, which immediately requires $\approx 10^{12}$ qubit machines to hope to match current classical results. 

\subsubsection{Numerical results}

We report numerical results to compare our \Algo{} algorithm to other baselines. We use the classical heuristic~\cite{Vogelstein2015} as before as a classical baseline, and the variational algorithm of~\cite{Mariella2023} as a quantum baseline. In the latter, we substitute their ansatz for $P$ with our ansatze for a fairer comparison. 

The graph isomorphism problem can be seen as a variant of a QAP~\cite{Vogelstein2015}. In fact,  
\[ \begin{aligned} 
\lVert A - P B P^\top \rVert^2_{\mathrm{F}} & = {\rm tr}((A - PBP^\top)^\top (A - PBP^\top)) \\
& = {\rm tr}( (A - PBP^\top) (A - PBP^\top)) 
\end{aligned} \]
by symmetry of the adjacency matrices \(A\) and \(B\), next
\[ \begin{aligned} 
\lVert A - P B P^\top \rVert^2_{\mathrm{F}} & = {\rm tr} ( A^\top A - A^\top P B P^\top - P B P^\top A + P B^\top P^\top P B P^\top) \\
& = {\rm tr}( A^\top A ) - {\rm tr}( A^\top P B P^\top ) - {\rm tr}( P B P^\top A ) + {\rm tr}(PB^\top B P^\top). 
\end{aligned} \]
We use the symmetry and the linearity of the trace operator to conclude with 
\[ \begin{aligned} 
\lVert A - P B P^\top \rVert^2_{\mathrm{F}} & = {\rm tr}(A^\top A) + 2 {\rm tr}(A P (-B) P^\top) + {\rm tr}(B^\top B) \simeq {\rm tr}(A P (-B) P^\top). 
\end{aligned} \]
Where $\simeq$ here stands for equivalence when optimized over. The latter is a QAP with $W = A$ and $D = -B$.

We generate graphs $\mathcal{A}$ randomly by sampling a Bernoulli distribution with a certain probability of $1$ set to $0.5$. The corresponding graph $\mathcal{B}$ is generated by drawing a random permutation and applying it to $\mathcal{A}$. The settings of the different algorithms is the same as in the QAP results. The step size of \ADAM{} is adjusted to $0.4$. 

\begin{figure}[H]
\centering
\includegraphics[width=0.95\textwidth]{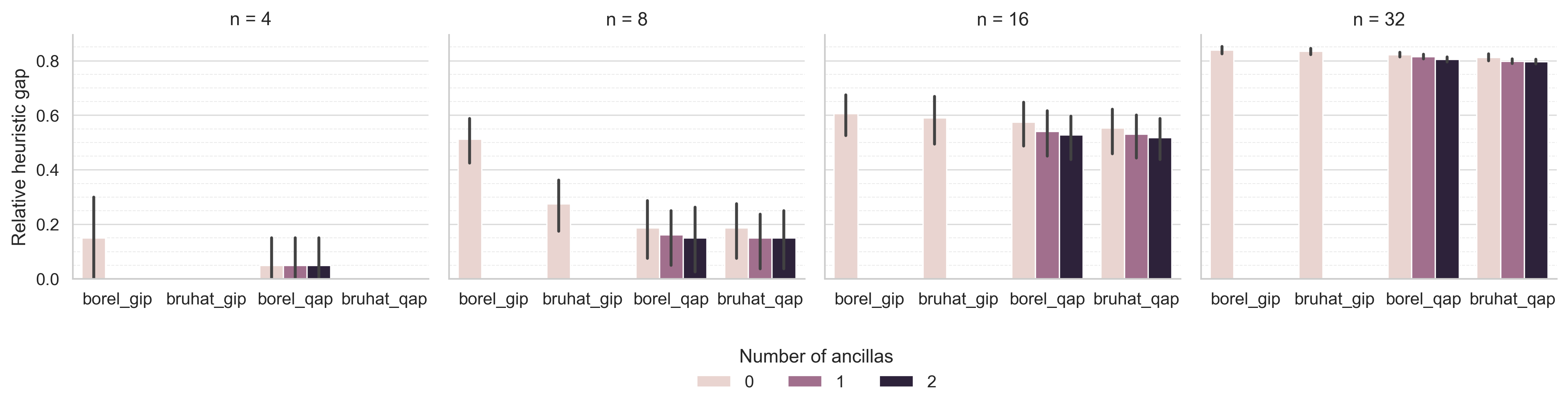}
\caption{Numerical comparisons on $10$ random instances of the graph isomorphism problem, with respect to the classical heuristic~\cite{Vogelstein2015}. We present here the mean over the instances and the 95\% confidence interval. The labels `borel\_gip' and `bruhat\_gip' represent the approach of~\cite{Mariella2023} with these ansatze, while `borel\_qap' and `bruhat\_qap' represent the new circuits introduced in this paper. }
\label{fig.gip}
\end{figure}

In Figure~\ref{fig.gip}, we report the numerical results for $n\in\{4,6,8,12,16\}$ comparing the variational quantum approach of~\cite{Mariella2023} where we have used our ansatze: Borel$^{ms}$ and Bruhat$^{ms}$, for their $P$, and our \Algo{} with Borel and Bruhat with different ancillas $m$. The comparison is with a normalized version of the heuristic gap, i.e., 
$$
\textrm{normalized heuristic gap} = \frac{\textrm{quantum result} - \textrm{heuristic classical result of~\cite{Vogelstein2015}}}{n^2/2}.
$$ 
The normalization is done with respect to the worst case error (and not with respect to the heuristic classical result, since it is often $0$). 

By Figure~\ref{fig.gip}, one can appreciate how our \Algo{} is competitive with respect the classical approach, especially for small instances, and it is better than the approach of~\cite{Mariella2023}. In larger instances, \Algo{} is less competitive, but as before, we think this is mainly due to the difficulties of the classical optimizer \ADAM{} to solve non-convex problems.  

\begin{figure}[ht]
\centering
\includegraphics[width=0.7\textwidth]{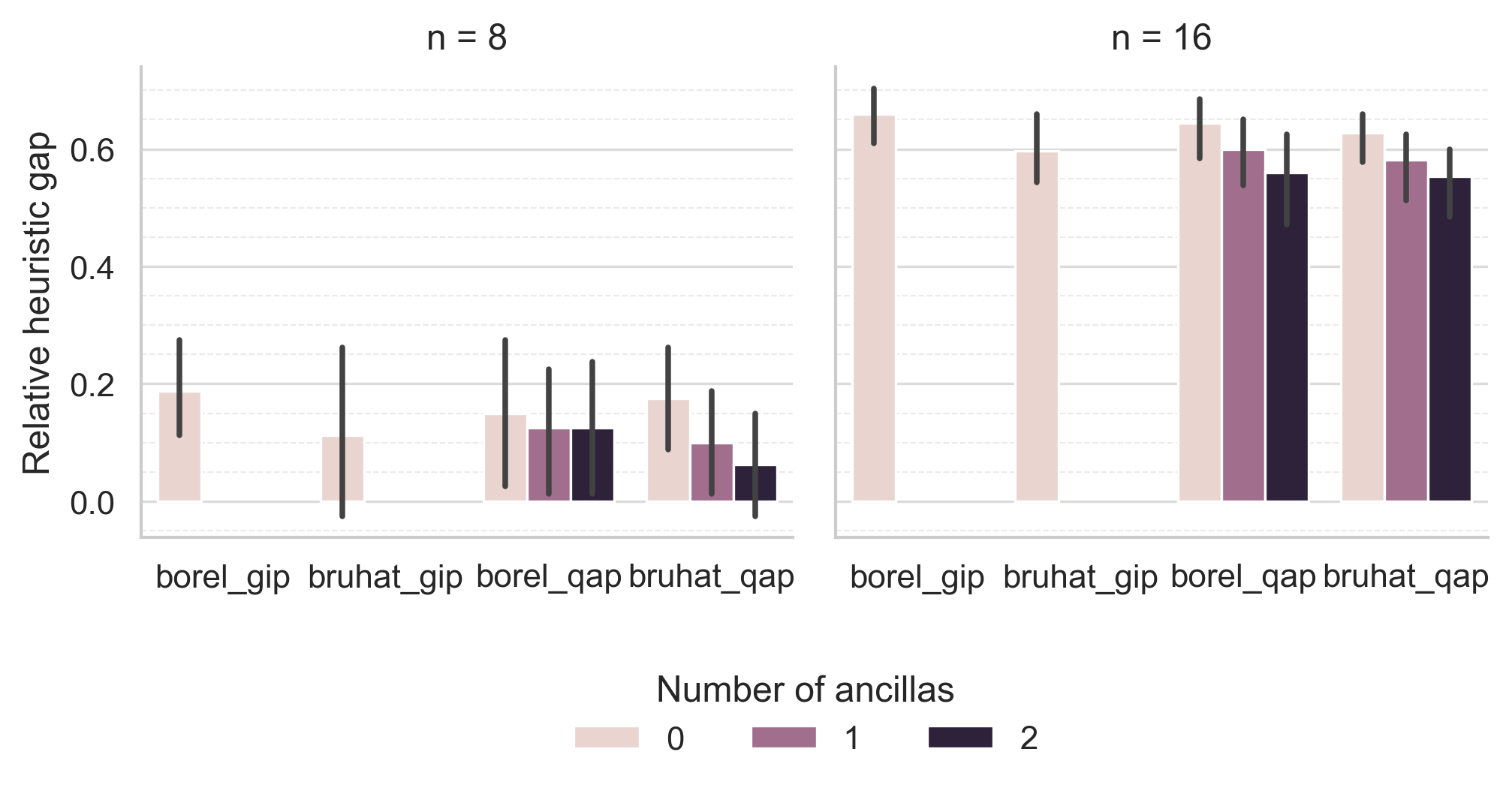}
\caption{Numerical comparisons on $10$ random instances of the graph isomorphism problem when the permutation lies in the span of Bruhat. }
\label{fig.gip.2}
\end{figure}

We then conclude our simulations testing the case in which the graph $\mathcal{B}$ is generated by a permutation which is in the span of Bruhat. The results are presented in Figure~\ref{fig.gip.2}, which shows better results than a general permutation, further advocating the need for an optimal spanning ansatz.

\section{Conclusions}

We have presented a variational quantum method to solve permutation-based optimization problems. The method is based on an optimal ansatz, with ancilla qubits, that can generate doubly-stochastic matrices. The resulting circuit is tunable and requires at least a number of qubits that scales only logarithmically with the number of variables. We have discussed and developed the theory behind the circuit generation and shown numerically its competitiveness on several optimization problems. We demonstrate that we can be in par with classical heuristics (typically for small instances but not only), and perform better than other quantum algorithms (on simpler problems, where those algorithms exist). The circuit that we generate is very general and adapted to any permutation-based optimization problems, which we will set forth to investigate in future work.

\bibliographystyle{quantum}
\bibliography{biblio}

\newpage
\appendix

\begin{center}
{\Large \sf Appendices}
\end{center}

\section{Computation complexity considerations}\label{ap.1}

We discuss briefly the computational complexity of~\cite{Rancic2023}. The variational algorithm is an amplitude encoding, requiring $q=\log_2(n)$ qubit to encode a max-cut of $n$ variables. However to compute the observable, one needs to evaluate $4^q = n^2$ circuits, weighted them with some pre-compute coefficient (involving the trace of a matrix-matrix multiplication) and then sum them. The total cost is $O(n^3)$. In comparison, a classical algorithm that evaluates $x^\top Q x$ requires only $O(n^2)$ floating point operations, thereby making the algorithm classically simulable.

\section{Construction of parametrized $\mathtt{cx}$ gate}\label{ap.2}

The reason why we cannot use the controlled rotation \texttt{x}, or \texttt{crx}, gate in our circuit is that it introduces some relative phase shift between the control qubit and the target qubit. To overcome this, we beforehand apply a phase gate on the control qubit, to anticipate the introduction of the relative phase with the \texttt{crx} gate. These two gates can be written as 
\[
\mathtt{p}(\theta) = \dyad{0}{0} + e^{i\theta} \dyad{1}{1} \qquad \text{and} \qquad \mathtt{crx}(\theta) = \dyad{0}{0} \otimes \mathtt{id} + \dyad{1}{1} \otimes \mathtt{rx}(\theta). 
\]
When evaluated at \(\theta = 0\), the \texttt{crx} gate does not introduce any relative phase as it acts like the identity, but when it is evaluated at \(\theta = \pi\) it becomes
\[
\mathtt{crx}(\pi) = \dyad{0}{0} \otimes \mathtt{id} + \dyad{1}{1} \otimes \mathtt{rx}(\pi) = \dyad{0}{0} \otimes \mathtt{id} - i \dyad{1}{1} \otimes \mathtt{x}. 
\]
We fix this issue by applying the phase gate on the control qubit with half of \(\theta\). 
\begin{figure}[h]
    \begin{center}
    \begin{tikzcd}[row sep = 0.2cm, column sep = 0.2cm]
        \ghost{\mathtt{p} \left( \phi/2 \right) } & \ctrl{1} & \qw \\
        \qw & \gate{\theta} & \qw 
    \end{tikzcd} = \begin{tikzcd}[row sep = 0.2cm, column sep = 0.2cm]
        \qw & \gate{\mathtt{p} \left( \theta/2 \right) } & \ctrl{1} & \qw \\
        \qw & \qw & \gate{\mathtt{rx}\left( \theta \right)} & \qw 
    \end{tikzcd}
    \end{center}
    \caption{Construction of the \texttt{crx} gate introducing no relative phase, using the notation of Figure~\ref{fig: first-borel}. }
    \label{fig: crxp}
\end{figure}
We have claimed that this gate acts like the identity when \(\theta = 0\), which is straightforward, and that it acts like a \texttt{cx} gate when \(\theta = \pi\). We prove the latter by direct computation. We denote by \(k\) the control qubit and by \(l\) the target qubit. 
\[ \begin{aligned} 
\mathtt{crx}_{kl}(\pi) \cdot \mathtt{p}_k(\pi/2) & = \Big( \dyad{0}{0}_k \otimes \mathtt{id}_l + \dyad{1}{1}_k \otimes \mathtt{rx}_l(\pi) \Big) \Big( \left( \dyad{0}{0}_k + e^{i \frac{\pi}{2}} \dyad{1}{1}_k \right) \otimes \mathtt{id}_l \Big) \\
& = \Big( \dyad{0}{0}_k \otimes \mathtt{id}_l - i \dyad{1}{1}_k \otimes \mathtt{x}_l \Big) \Big( \left( \dyad{0}{0}_k + i \dyad{1}{1}_k \right) \otimes \mathtt{id}_l \Big) \\
& = \dyad{0}{0}_k \otimes \mathtt{id}_l - i \cdot i \dyad{1}{1}_k \otimes \mathtt{x}_l \\
& = \mathtt{cx}_{kl}. 
\end{aligned} \]

\section{Projection onto permutations}\label{ap.permutation}

Let $\hat{P}$ be a doubly-stochastic matrix of dimension $n$, i.e., a matrix in $[0,1]^{n \times n}$, whose columns and rows sum to $1$. Projecting such a matrix onto the permutation set, we can either use a linear assignment problem (a convex problem) as,
$$
\min_{P \in \Pi_n}\, \frac{1}{2} \lVert \hat{P} - P \rVert^2_{\mathrm{F}} = - \sum_{i=1}^{n} \sum_{j=1}^n (\hat{P}^\top P)_{ij}, 
$$
that we solve using the \emph{Hungarian algorithm}, or a randomized algorithm described in~\cite{Barvinok2006,Fogel2015}. We explain here the latter as it has in general a better performance on small instances and is less known. 

The randomized algorithm idea is apply $\hat{P}$ to a sorted random vector and see how the ordering changes. Then one can find a permutation that achieve the same permutation of ordering. Let $v$ the initial random vector, and $\psi(v)$ its ordering, the idea is to find the permutation $P$ for which,
$$
\psi \left( \hat{P} v \right) = P \psi(v).
$$ 

In the simulations, we choose to start with a random vector $v$ designed as \(v_i \coloneqq 2^i\). Each coefficient of \(v\) is at least twice or half another coefficient, which prevents values of \(\hat{P}\) which are quite far from \(1\) to unexpectedly change the order of the returned vector. 

Next, we randomly permute \(v\), apply \(\hat{P}\) to it, and construct \(P\) by looking at the order of the returned vector. We do this \(50\) times, so we obtain less than \(50\) unique permutations, closely resembling the one we are looking for. From these, we simply take the one minimizing the objective function. 

This approach is efficient, and allows for the exploration of a larger neighborhood of the current relaxed solution, compared to the projection that we can have by solving a linear assignment problem, which does not take into consideration the cost function.

\section{Full results for QAP}\label{ap.qapres}

The following figures show the average relative optimality gaps for the different algorithms. One plot corresponds to a given size of instances. For the size \(128\) and \(256\), there is a unique instance, so there is no vertical bar indicating the interval of confidence.


The following tables includes all the results we obtained on the QAPlib instances of size a power of two. Their names are written in the first column. So each line represents a single instance, and the columns represent the different algorithms used to solve the aforementioned instances.

\begin{figure}[H]
\begin{center}
\begin{adjustbox}{width=\textwidth}
\begin{tabular}{@{}lcccccccccccc@{}} \toprule
 & \multicolumn{3}{c}{Classical} & \multicolumn{3}{c}{Borel} & \multicolumn{3}{c}{Bruhat} & \multicolumn{3}{c}{SEL} \\ \cmidrule(lr){2-4} \cmidrule(lr){5-7} \cmidrule(lr){8-10} \cmidrule(l){11-13}
Instances & Exact & Heur. & Rand. & \(0\) & \(1\) & \(2\) & \(0\) & \(1\) & \(2\) & \(0\) & \(1\) & \(2\) \\ \midrule
esc16a     & 68      & 70     & 84       & 84 & 80 & 80 & 80 & 78 & 78 & 84 & 78 & 78 \\
esc16b     & 292    & 320   & 292     & 294 & 294 & 294 & 296 & 292 & 292 & 294 & 294 & 294 \\
esc16c     & 160    & 168   & 192     & 178 & 178 & 168 & 182 & 182 & 182 & 188 & 184 &180  \\
esc16d     & 16     & 62     & 26        & 24 & 24 & 16 & 28 & 16 & 16 & 28 & 24 & 24 \\
esc16e     & 28     & 30     & 30        & 32 & 32 & 32 & 32 & 32 & 32 & 34 & 34 & 32 \\
esc16f      & 0       & 0       & 0          & 0 & 0 & 0 & 0 & 0 & 0 & 0 & 0 & 0 \\
esc16g     & 26     & 30     & 36        & 30 & 30 & 28 & 34 & 34 & 34 & 36 & 36 & 36 \\
esc16h     & 996   & 1518 & 1066    & 1026 & 1026 & 1026 & 1026 & 1026 & 1026 & 1036 & 1030 & 1030 \\
esc16i      & 14     & 14     & 26        & 18 & 18 & 14 & 22 & 22 & 20 & 18 & 14 & 14 \\
esc16j      & 8       & 8       & 12        & 12 & 12 & 10 & 12 & 12 & 12 & 12 & 10 & 10 \\
nug16a    & 1610 & 1660 & 1916    & 1860 & 1860 & 1854 & 1878 & 1778 & 1778 & 1866 & 1866 & 1844 \\
nug16b    & 1240 & 1282 & 1516    & 1486 & 1446 & 1446 & 1434 & 1434 & 1434 & 1488 & 1484 & 1484 \\ \bottomrule
\end{tabular}
\end{adjustbox}
\caption{The value of the cost function on different instances of size 16. The integers below ansatze names indicate the number \(m\) of ancilla qubits used.}
\end{center}
\end{figure}

\begin{figure}[H]
\begin{center}
\begin{adjustbox}{width=\textwidth}
\begin{tabular}{@{}lcccccccccccc@{}} \toprule
 & \multicolumn{3}{c}{Classical} & \multicolumn{3}{c}{Borel} & \multicolumn{3}{c}{Bruhat} & \multicolumn{3}{c}{SEL} \\ \cmidrule(lr){2-4} \cmidrule(lr){5-7} \cmidrule(lr){8-10} \cmidrule(l){11-13}
Instances & Exact & Heur. & Rand. & \(0\) & \(1\) & \(2\) & \(0\) & \(1\) & \(2\) & \(0\) & \(1\) & \(2\) \\ \midrule
kra32 & \footnotesize{88900} & \footnotesize{92930} & \footnotesize{126020} & \footnotesize{120710} & \footnotesize{118710} & \footnotesize{115710} & \footnotesize{117600} & \footnotesize{114680} & \footnotesize{114680} & \footnotesize{122390} & \footnotesize{122390} & \footnotesize{122390} \\
esc32a & 130 & 160 & 338 & 318 & 304 & 304 & 282 & 258 & 258 & 306 & 306 & 306 \\
esc32b & 168 & 196 & 376 & 264 & 256 & 256 & 256 & 256 & 256 & 352 & 304 & 304 \\
esc32c & 642 & 650 & 810 & 702 & 694 & 694 & 706 & 692 & 692 & 776 & 742 & 742 \\
esc32d & 200 & 226 & 294 & 262 & 262 & 262 & 262 & 258 & 254 & 280 & 280 & 272 \\
esc32e & 2 & 2 & 6 & 6 & 6 & 6 & 6 & 6 & 2 & 6 & 6 & 6 \\
esc32g & 6 & 10 & 18 & 12 & 10 & 10 & 12 & 10 & 10 & 8 & 8 & 8 \\ \bottomrule
\end{tabular}
\end{adjustbox}
\caption{The value of the cost function on different instances of size 32. The integers below ansatze names indicate the number \(m\) of ancilla qubits used.}
\end{center}
\end{figure}

\begin{figure}[H]
\begin{center}
\begin{adjustbox}{width=\textwidth}
\begin{tabular}{@{}lcccccccccccc@{}} \toprule
 & \multicolumn{3}{c}{Classical} & \multicolumn{3}{c}{Borel} & \multicolumn{3}{c}{Bruhat} & \multicolumn{3}{c}{SEL} \\ \cmidrule(lr){2-4} \cmidrule(lr){5-7} \cmidrule(lr){8-10} \cmidrule(l){11-13}
Inst. & Exact & Heur. & Rand. & \(0\) & \(1\) & \(2\) & \(0\) & \(1\) & \(2\) & \(0\) & \(1\) & \(2\) \\ \midrule
sko64 & 48498 & 49102 & 56450 & 56564 & 56482 & 56482 & 56120 & 55592 & 55592 & 56880 & 56612 & 56582 \\
tai64c & 1855928 & 5893540 & 2258166 & 2141126 & 2141126 & 2141126 & 2211470 & 2181600 & 2161392 & 2187554 & 2133174 & 2133174 \\ 
esc128 & 64 & 72 & 262 & 128 & 114 & 114 & 148 & 144 & 134 & 226 & 182 & 182 \\
tai256c & 44759294 & 98685678 & 50647386 & 50129154 & 50129154 & 50129154 & 50093960 & 49437762 & 49035944 & 50642900 & 50140242 & 50140242 \\ \bottomrule
\end{tabular}
\end{adjustbox}
\caption{The value of the cost function on different instances of size 64, 128 and 256. The integers below ansatze names indicate the number \(m\) of ancilla qubits used.}
\end{center}
\end{figure}

\begin{figure}[H]
\begin{center}
\includegraphics[width=0.95\textwidth]{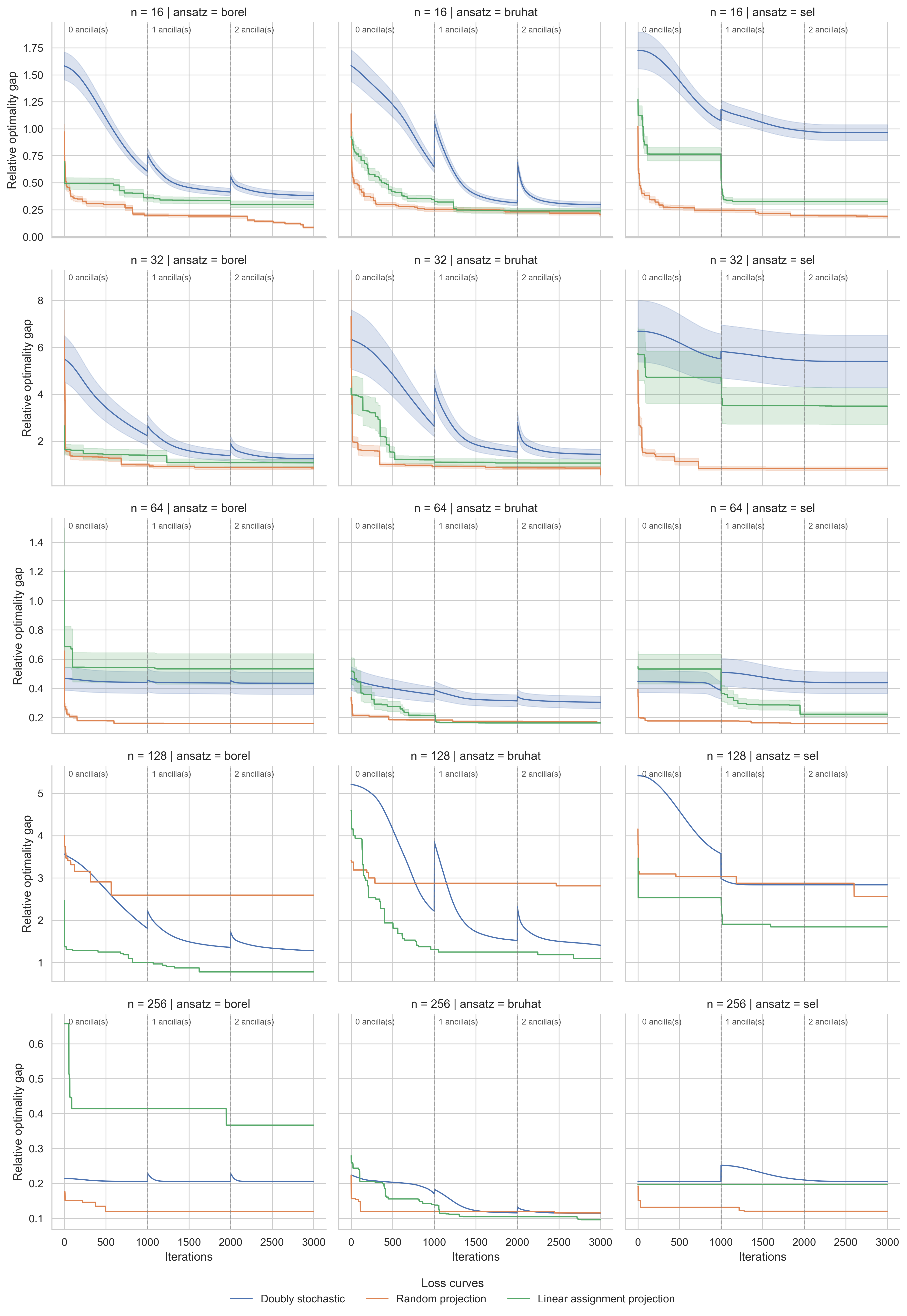}
\caption{Values of the cost functions during the optimization process. Each row correspond to a specific size of instances. For \(128\)- and \(256\)-vertex graphs, there is a unique instance. Vertical dashed lines indicate the iterations at which an additional ancilla qubit is introduced in the variational ansatz.}
\end{center}
\end{figure}

\end{document}